\pdfoutput=1
\documentclass[preprint,floats,aps,12pt,superscriptaddress,nofootinbib,floatfix]{revtex4}
\usepackage[utf8]{inputenc}
\usepackage{amsmath}
\usepackage{amsfonts}
\usepackage{amssymb}
\usepackage{bm}
\usepackage{natbib} 
\usepackage{times} 
\usepackage{prelim2e}\usepackage[none,bottom]{draftcopy}
\draftcopyName{preprint / }{1.2} %ADAPT TEXT
\usepackage[pdftex]{graphicx}
\usepackage[pdftex]{hyperref}
\hypersetup{
	pdftitle={Efficient calculation of phase coexistence and phase diagrams: application to a binary phase-field crystal model} %ADAPT TEXT 
	pdfauthor={Max Philipp Holl, Andrew J. Archer, and Uwe Thiele},  %ADAPT TEXT 
	pdfproducer={lateX},
	pdfview=FitV,       % FitH
	pdfstartview=FitB,
	linkcolor=blue,     % links to same page
	citecolor=blue,     % citations
	urlcolor=red,      % links to URLs
	breaklinks=true,    % links may be split onto 2 lines
	colorlinks=true,
	citebordercolor=0 0 0,  % color for \cite
	filebordercolor=0 0 0,
	linkbordercolor=0 0 0,
	menubordercolor=0 0 0,
	urlbordercolor=0 0 0,
	pdfhighlight=/I,
	pdfborder=0 0 0,   % no box around links
	bookmarksopen=true,
	bookmarksnumbered=true
}

\usepackage[usenames,dvipsnames]{color}
\usepackage[normalem]{ulem}

\newcommand{\dt}{\partial_t}

\newcommand{\bphi}{\bar{\phi}}

\renewcommand{\vec}[1]{\bm{#1}}

\graphicspath{{./figures/}}

\begin{document}
\title{Efficient calculation of phase coexistence and phase diagrams: application to a binary phase-field crystal model}
\author{Max Philipp Holl}
\email{m.p.holl@wwu.de}
\thanks{ORCID ID: 0000-0001-6451-9723}
\affiliation{Institut f\"ur Theoretische Physik, Westf\"alische Wilhelms-Universit\"at M\"unster, Wilhelm Klemm Str.\ 9, 48149 M\"unster, Germany}

\author{Andrew J. Archer}
\email{a.j.archer@lboro.ac.uk}
\thanks{ORCID ID: 0000-0002-4706-2204}
\affiliation{Department of Mathematical Sciences, Loughborough University, Loughborough LE11 3TU, United Kingdom}
\affiliation{Interdisciplinary Centre for Mathematical Modelling, Loughborough University, Loughborough LE11 3TU, United Kingdom}

\author{Uwe Thiele}
\email{u.thiele@uni-muenster.de}
\homepage{http://www.uwethiele.de}
\thanks{ORCID ID: 0000-0001-7989-9271}
\affiliation{Institut f\"ur Theoretische Physik, Westf\"alische Wilhelms-Universit\"at M\"unster, Wilhelm Klemm Str.\ 9, 48149 M\"unster, Germany}
\affiliation{Center of Nonlinear Science (CeNoS), Westf{\"a}lische Wilhelms-Universit\"at M\"unster, Corrensstr.\ 2, 48149 M\"unster, Germany}
\affiliation{Center for Multiscale Theory and Computation (CMTC), Westf{\"a}lische Wilhelms-Universit\"at, Corrensstr.\ 40, 48149 M\"unster, Germany}

\begin{abstract}
We show that one can employ well-established numerical continuation methods to efficiently calculate the phase diagram for thermodynamic systems. In particular, this involves the determination of lines of phase coexistence related to first order phase transitions and the continuation of triple points. To illustrate the method we apply it to a binary Phase-Field-Crystal model for the crystallisation of a mixture of two types of particles. The resulting phase diagram is determined for one- and two-dimensional domains. In the former case it is compared to the diagram obtained from a one-mode approximation. The various observed liquid and crystalline phases and their stable and metastable coexistence are discussed as well as the temperature-dependence of the phase diagrams. This includes the (dis)appearance of critical points and triple points. We also relate bifurcation diagrams for finite-size systems to the thermodynamics of phase transitions in the infinite-size limit.
\end{abstract}
\maketitle
%\newpage
%\tableofcontents
%\newpage

%\clearpage

%\ttuwe{We need some more citations for the ``phase-diagram literature'', in general, at some points some references could be added; also citations in intro and conclusions might not be consistent}

\section{Introduction}\label{sec:introduction}

We introduce an efficient method to calculate the phase behaviour of thermodynamic systems. To demonstrate the method, we apply it to a two-component Phase Field Crystal (PFC) model, which describes the crystallisation behaviour of a binary mixture of (colloidal) particles. Continuum models of this kind are widely used to describe the liquid to crystalline solid phase transition \cite{ELWG2012ap}. PFC models are in many regards similar to other families of models, such as those arising from Density Functional Theory (DFT), from which they can be derived (cf.~\cite{ARRS2019pre, ELWG2012ap}). The techniques described here can be applied to these models as well.

PFC and PFC-type models are increasingly widely used. A two-dimensional model similar to the one we consider here was used to study pattern formation in lipid bilayers \cite{HiKA2009c}. More recently, the phase behaviour of a closely related model of binary colloidal crystals was presented by Taha et al.~\cite{TDME2019prm}, also for the two-dimensional case. As well as for modelling the phase behaviour of matter, PFC-type models for biological and active systems have been developed \cite{ophaus2018resting} and even for quantum-mechanical systems, such as in the work of Heinonen et al.\ who developed a PFC-like model for superfluidity and supersolidity in Bose-Einstein condensates \cite{HeBD2019pra}.

In Ref.~\cite{HiKA2009c}, to determine the phase diagram a one-mode approximation was used, i.e.\ it is assumed that the crystalline state may be approximated by harmonics of a single wavelength. This approach is similar to that used in the derivation of amplitude equations in the area of pattern formation \cite{CrHo1993rmp,Hoyle2006}. Similarly, a single-mode approach together with direct numerical calculations was used to determine the structures and phase diagrams in the work of Taha et al.~\cite{TDME2019prm}. Tracking the coexistence curves between the different phases using direct numerical calculations can be time consuming, which is why one-mode type approximations are often used. However, comparing the phase diagram of the one-component PFC model calculated using a one-mode approximation (c.f.~Ref.~\cite{ElGr2004pre}) to the exact phase diagram obtained from a full numerical solution, the limitations of the one-mode approximation become apparent even in one dimension \cite{TARG2013pre}. In particular, there it is shown that more than one mode is needed to exactly obtain the tricritical point where the phase transition changes from being first order to second order. For two-dimensional systems, Ref.~\cite{TDME2019prm} shows examples where there is relatively poor agreement between the phase diagram obtained via a one-mode approximation and that from direct numerical simulations. In particular, problems arise in regions of two- and three-phase coexistence. Moreover, it seems some phase boundaries could not be completely obtained with their methods (see their Fig.~3). 

Additionally, deriving the amplitude equations can be cumbersome in systems with different length scales. In these more complex cases one mode is generally not sufficient and two or more modes need to be considered. In contrast, our fully nonlinear approach based on the well established numerical path continuation methods that are widely used in pattern formation problems and the study of complex systems \cite{KrauskopfOsingaGalan-Vioque2007,DWCD2014ccp,EGUW2019springer} enables us to track easily the coexistence phase boundaries between the different phases. Code packages such as \textsc{auto07p} \cite{DoKK1991ijbc,DoedelOldeman2009} and \textsc{pde2path} \cite{UeWR2014nmtmaa} provide all the routines required. These methods have recently been applied to study the emergence of the Maxwell construction for phase coexistence in both the Cahn-Hilliard and the PFC model \cite{TFEK2019njp}. Once set up to study the bifurcation diagrams in finite size systems, these continuation code can easily be extended to numerically calculate the phase diagrams. Of course, a point that merits careful consideration is the choice of the numerical domain for the crystal phases, since it has to have the right size and boundary conditions to accommodate the crystalline structures. These are, however, general considerations that also apply in most other approaches.

This paper is structured as follows: In section~\ref{sec:binary-pfc} we introduce the two-field PFC model used to illustrate the proposed method. We discuss some general considerations for both the one-mode approximation and the fully nonlinear solution method in section~\ref{sec:general_considerations}. In the subsequent section~\ref{sec:one-mode} we present the simplified approach, i.e.\ the one-mode approximation for our system. Section~\ref{sec:coexistence-continuation} explains our method in more detail while section~\ref{sec:results} presents selected results for a one-dimensional (1d) system and compares them to results obtained employing a one-mode approximation. In section~\ref{sec:triple-points} we extend the method to follow triple points in parameter space while in section~\ref{sec:2D_results} results are presented for a two-dimensional (2d) system. Finally, in section \ref{sec:conclusions} we conclude and discuss further investigations facilitated by the proposed methods. 

\section{Two-field Phase Field Crystal model}\label{sec:binary-pfc}

We calculate the fully nonlinear phase behaviour of a two-field PFC model that is derived from the Helmholtz free energy:
\begin{align}
          \mathcal{F}[\vec{\phi}] = \int_V\text{d}^nx \left[\sum_{j = 1}^{2} \left\lbrace \frac{\phi_j}{2}[r + (q_j^2 + \Delta)^2]\phi_j + \frac{\phi_j^4}{4}\right\rbrace	+ c\phi_1\phi_2\right].
          \label{eq:cPFC3_FE}
\end{align}
        where $V$ is the ``volume'' of the considered $n$-dimensional domain.
Further, the $ \phi_j $ with $j=1,2$ are order parameter fields associated with the two different species of particles. In the derivation of PFC models from DFT they arise as scaled densities \cite{ARRS2019pre, ELWG2012ap}. For simplicity, here we refer to these as concentrations. The parameter $ r $ is a scaled shifted temperature, sometimes called the ``undercooling''. The $ q_j $ are the critical wavenumbers for the two decoupled fields, i.e.\ they control the length scale of the crystalline patterns formed, $ c $ is a coupling constant and $ \Delta $ denotes the Laplacian. In principle it is possible for the two species to have different freezing temperatures, which can be incorporated by replacing $ r $ with $ r + \Delta r $ in one of the two terms of the sum in \eqref{eq:cPFC3_FE}, with $ \Delta r $ then being the difference in the scaled temperatures. However, here we do not consider this case and, therefore, keep the above notation. Further, we only consider the case $ q_1=q_2=1 $ and correspondingly drop the subscript $j$ where convenient. This free energy reduces to the standard PFC model free energy \cite{ElGr2004pre,TARG2013pre} for each field if one or other of the concentrations becomes zero everywhere. Additionally, there is the single coupling term $ c\phi_1\phi_2 $ that represents the simplest form of interaction energy. It keeps the $ \phi_j \rightarrow -\phi_j $ symmetry of the decoupled systems.

Using the free energy~\eqref{eq:cPFC3_FE}, the dynamics is given by the conserved gradient dynamics \cite{ELWG2012ap}
\begin{align}
%          \dt \vec{\phi} &=
                           \dt\left(\begin{array}{c}
\phi_1\\\phi_2
\end{array}\right) &= \Delta\left(\begin{array}{c}
\frac{\delta \mathcal{F}}{\delta\phi_1} \\\frac{\delta \mathcal{F}}{\delta\phi_2}
\end{array}\right) \nonumber\\&= \Delta\left(\begin{array}{c}
[r + (q_1^2 + \Delta)^2]\phi_1 + \phi_1^3 + c\phi_2\\
\left[r + (q_2^2 + \Delta)^2\right]\phi_2 + \phi_2^3 + c\phi_1
\end{array}\right),\label{eq:cPFC3}
\end{align}
        where $t$ is time and where ``conserved'' refers to the fact that this dynamics conserves the masses $m_j=\int_V\text{d}^nx\phi_j$ (and the mean densities $\bar\phi_j=m_j/V$). Note that we use identical mobilities in both equations that are then absorbed into the time scale. This simplification does not affect the phase behaviour, but needs to be justified when studying dynamic behaviour.
Overall, Eqs.~\eqref{eq:cPFC3} are parity and field-inversion symmetric, i.e., they do not change their form for $\vec{x}\to-\vec{x}$ and $(\phi_1,\phi_2)\to-(\phi_1, \phi_2)$, respectively. For $q_1=q_2$ the symmetry w.r.t.\ exchange of the two fields $(\phi_1,\phi_2)\to (\phi_2, \phi_1)$ also holds.
        Our attention is focused on steady states, i.e., $ \dt \phi_j = 0 $. With this, Eqs.~\eqref{eq:cPFC3} can be integrated twice to obtain 
\begin{align}
\left(\begin{array}{c} 0\\ 0 \end{array}\right)
= \left(\begin{array}{c}
[r + (q_1^2 + \Delta)^2]\phi_1 + \phi_1^3 + c\phi_2 - \mu_1\\
\left[r + (q_2^2 + \Delta)^2\right]\phi_2 + \phi_2^3 + c\phi_1 - \mu_2
\end{array}\right)\label{eq:twice_integrated}
\end{align}
where the chemical potentials $ \mu_j$ arise as integration constants. Note that the integration constants of the first integration are set to zero as we exclude fluxes across the boundaries.

        When considering steady states on finite size domains we present them in state or bifurcation diagrams that display typical measures that characterize the states as a function of a control parameter such as, e.g., a mean concentration or a chemical potential. As characterizing measures, often the L$ _2 $-norm of the deviation of the concentrations from their mean values
\begin{align}
||\delta\vec{\phi}|| = \sqrt{\frac{1}{L}\int_{-L/2}^{L/2}\text{d}x [(\phi_1 - \bar{\phi}_1)^2 + (\phi_2 - \bar{\phi}_2)^2] }
\end{align}
is used, as well as the mean Helmholtz free energy density
\begin{align}
\bar{f} = \frac{\mathcal{F}}{V}, \label{eq:cPFC3_FED}
\end{align}
      and the mean grand potential density
% \begin{align}
% \Omega = \mathcal{F} - \int \text{d}^nx(\mu_1\phi_1 + \mu_2\phi_2)
% \end{align}
% and its 
\begin{align}
          \bar{\omega} = \bar{f} - \mu_1\bar{\phi}_1 - \mu_2\bar{\phi}_2.
          \label{eq:gpd}
\end{align}

Before considering the phase behaviour in the following section~\ref{sec:calccoex}, here we first briefly describe the essential ideas in path continuation methods \cite{KrauskopfOsingaGalan-Vioque2007, DWCD2014ccp, EGUW2019springer} and also analyse the linear stability of homogeneous (i.e., liquid) states of different concentrations and give selected examples of bifurcation diagrams. The basic underlying concept in path continuation methods is as follows: Consider some system of differential equations 
\begin{align}
	 \dt u = G(u, \lambda), \label{eq:diff_eq}
\end{align}
with variables $ u $ and control parameter(s) $ \lambda $. Note that once a suitable spatial discretization has been applied, partial differential equations such as the system in Eq.~\eqref{eq:cPFC3} have the form \eqref{eq:diff_eq}. The main idea is to follow a branch of steady state solutions, i.e., states satisfying $ \dt u=0 $, through parameter space. To obtain a bifurcation diagram, one starts from a known solution $ u_0 $, e.g., the homogeneous state. From there, the continuation parameter $ \lambda $ is varied by a certain stepsize and a new solution is found by solving Eq.~\eqref{eq:diff_eq} at the new parameter value using $ u_0 $ as an initial guess. The stability of these solutions can be determined, e.g., by calculating the eigenvalues of the Jacobian of $ G(u, \lambda) $. A change in the number of unstable eigenvalues then indicates a bifurcation. Existing code packages contain routines to follow a solution branch, to detect bifurcations, and switch to a branch emerging at a bifurcation. As described, this continuation routine using the \textit{natural parametrization} fails at points where a solution branch folds back, i.e., at saddle-node bifurcations. To avoid this, one alternatively employs \emph{pseudo-arclength parametrization}, where the arclength $s$ of the solution branch is used as the continuation parameter. Then, the control parameter $\lambda$ becomes part of the solution and the system of equations to be solved is extended by adding the pseudo-arclength equation
\begin{align}
	\dot{u}(u - u_0) + \dot{\lambda}(\lambda - \lambda_0) = \Delta s, \label{eq:pseudoarclength}
\end{align}
to the system of equations being solved, where $ (\dot{u}, \dot{\lambda}) $ is the direction of the branch and $ \Delta s $ is the step size in the continuation parameter along the solution branch. This allows continuation through fold points. 

\subsection{Linear stability analysis and typical bifurcation diagrams}

Linearising Eq.~\eqref{eq:cPFC3} about the homogeneous states $ (\phi_1, \phi_2)^T  = (\bar{\phi}_1, \bar{\phi}_2)^T $ and employing the Ansatz $ \delta\phi_j(x, t) \equiv \phi_j(x, t) - \bar\phi_j = \chi_j\exp(\sigma t + ikx)  $, with $(\chi_1,\chi_2)^T$ the eigenvector to an eigenvalue $ \sigma $ results in the dispersion relations 
\begin{align}
\sigma_\pm = -\frac{k^2}{2}\left(J_1 + J_2 \pm \sqrt{(J_1 - J_2)^2 + 4c^2}\right),
\end{align}
where $ J_i = r + (q_i^2 - k^2)^2 + 3\bar{\phi}_i^2 $. 
When the two intrinsic length scales are identical, i.e. $ q_1=q_2=q $, the relation simplifies to
\begin{align}
\sigma_\pm = -\frac{k^2}{2}\left\lbrace 2[r + (q^2-k^2)^2] + 3(\bar{\phi}_1^2 + \bar{\phi}_2^2) \pm \sqrt{9(\bar{\phi}_1^2 - \bar{\phi}_2^2)^2 + 4c^2}\right\rbrace.
\end{align}
As expected for a gradient dynamics, the eigenvalues are always real. The onsets of the two modes of short-scale instability occur for $k=k_\mathrm{o}=q$ at $r=r_\mathrm{c}^\pm=q^2\,[-3(\bar{\phi}_1^2 + \bar{\phi}_2^2) \pm \sqrt{9(\bar{\phi}_1^2 - \bar{\phi}_2^2)^2 + 4c^2}]/2$. Decreasing the temperature $r$ from large values, the first mode to become unstable is the one with the ``$+$'' sign, i.e., for $\bar{\phi}_1=\bar{\phi}_2=0$ at $r_\mathrm{c}^+=|c| q^2$.

Above onset, i.e., for $r<r_\mathrm{c}^+$, a finite band of unstable wavenumbers exist. Whenever $\sigma_+$ or $\sigma_-$ crosses zero while changing one of the control parameters, a pitchfork bifurcation occurs. There, a branch of inhomogeneous steady states emerges from the branch of homogeneous states. For instance, when employing one of the mean densities $\bar{\phi}_j$ as the control parameter, fixing all remaining parameters and the domain size (that controls $k_\mathrm{o}$), a bifurcation occurs at critical values $\bar{\phi}_j^\mathrm{c}$. Here, we detect such bifurcations within the path continuation routines and follow the emerging branches of inhomogeneous states.

\begin{figure}
	\centering
	\includegraphics[width=\textwidth]{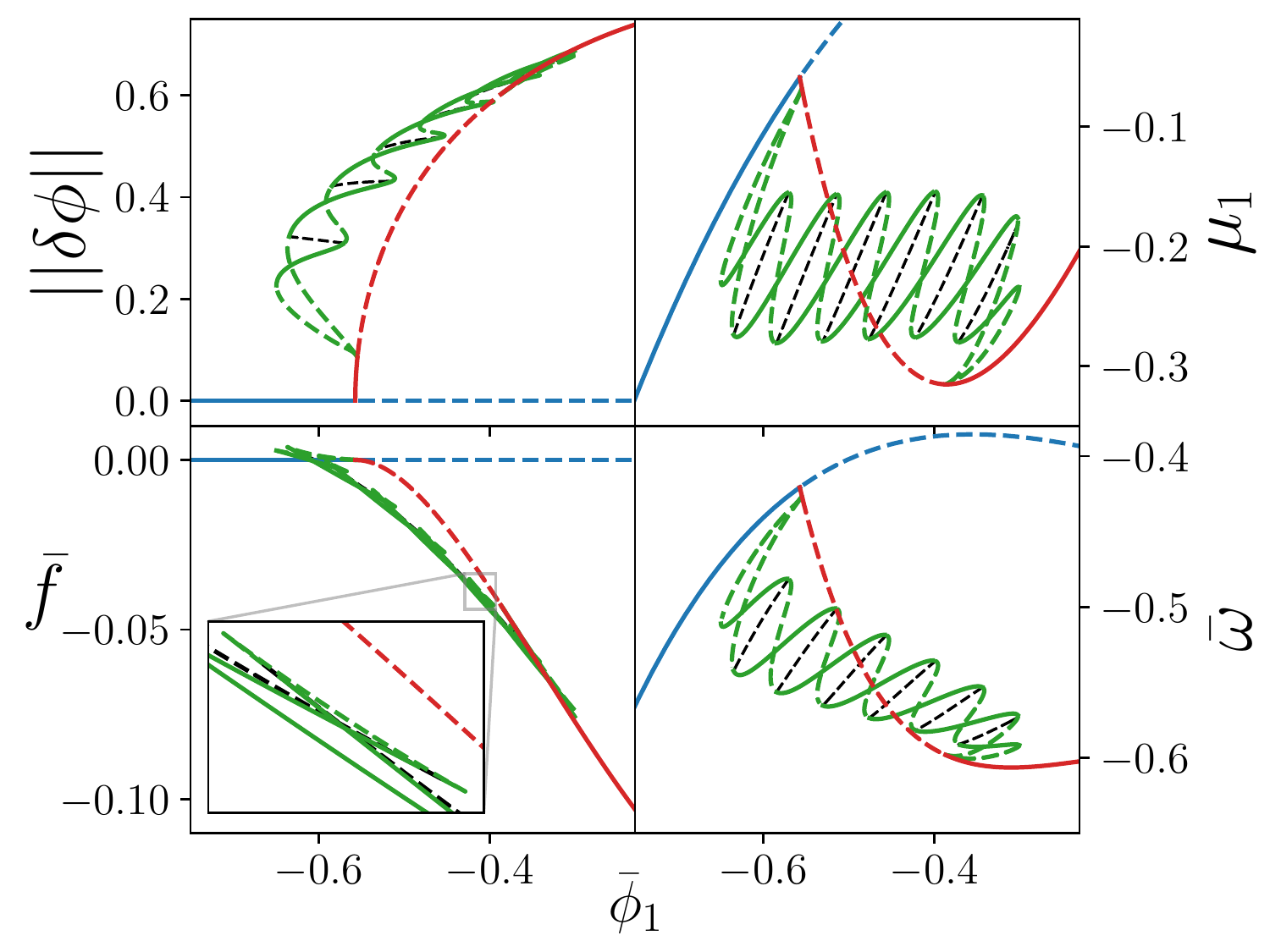}
	\caption{Typical bifurcation diagrams for the binary PFC in Eq.~\eqref{eq:cPFC3}, exhibiting four types of steady states: uniform liquid state (blue line), periodic crystal state (red line), symmetric (green lines) and asymmetric (black lines) localized states. The four panels show the same solution branches using (a) the norm, (b) the chemical potential $ \mu_1 $, (c) the mean Helmholtz free energy density, and (d) the mean grand potential density as solution measure. Stable [unstable] steady states are marked by solid [dashed] lines. The concentration $ \bar{\phi}_2 $ is fixed at $ \bphi_2 = -0.85 $. The other parameters are $ r = -0.9 $, $ q_1 = q_2 = 1 $, and $ c = -0.2 $. The domain size is $ L = 16\pi $.}
	\label{fig:bifdiag_general}
\end{figure}

   \begin{figure}
\centering
\includegraphics[width=\textwidth]{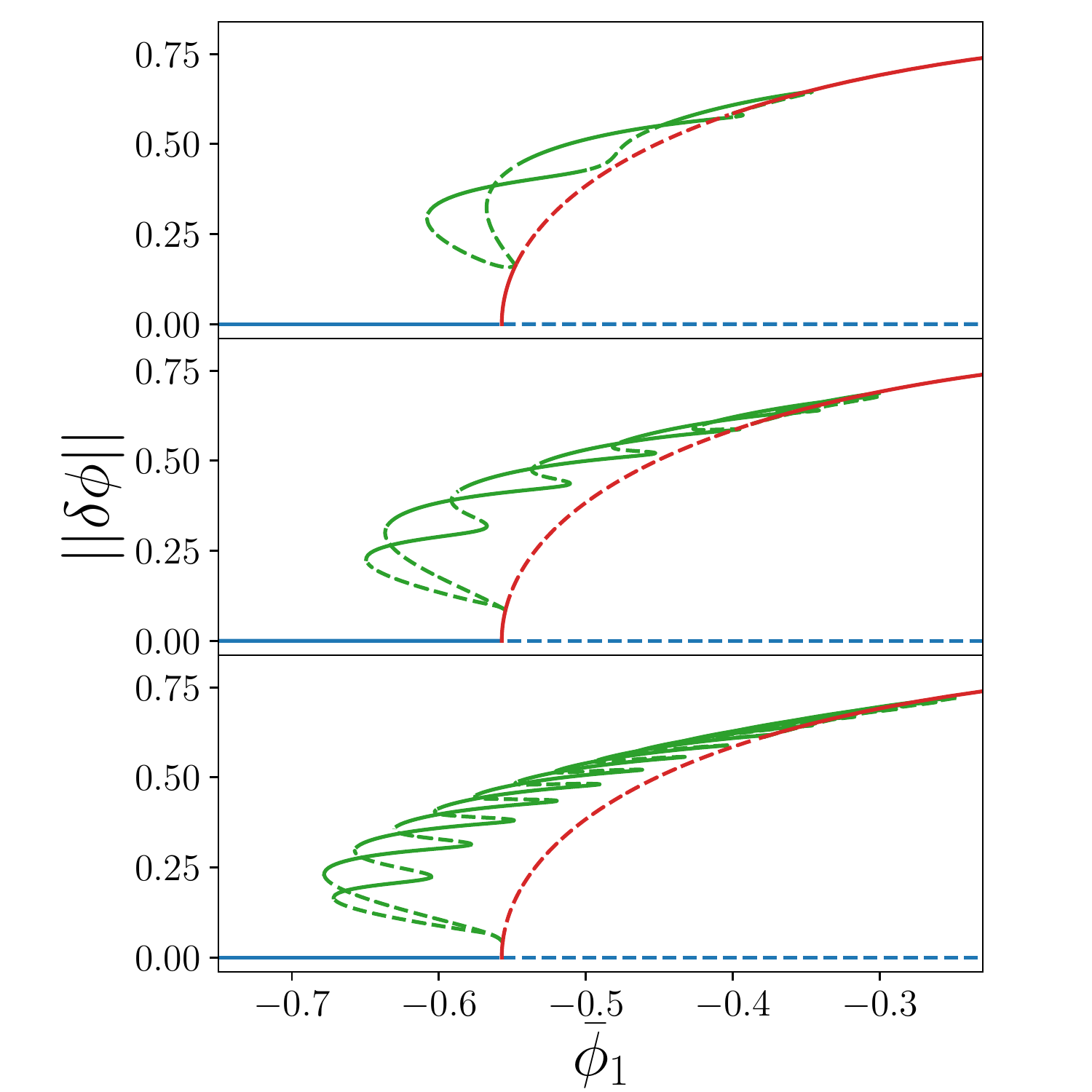}
\caption{Bifurcation diagrams of the binary PFC in Eq.~\eqref{eq:cPFC3}, displayed in terms of the mean grand potential as in Fig.~\ref{fig:bifdiag_general}~(d) are shown for three different domain sizes $ L=8\pi, 16\pi, 32\pi $, going from top to bottom. The remaining parameters and line styles are as in Fig.~\ref{fig:bifdiag_general}. For clarity, the asymmetric localized states are not included.
  % Since a pair of density peaks grows to full size on the unstable parts of a localized branch and more density peaks fit in a larger domain, the snaking appears more crowded for larger domains.}
}
\label{fig:bifdiag_domainSize}
\end{figure}

A typical example of the resulting bifurcation diagrams with $\bphi_1$ as the control parameter is displayed in Fig.~\ref{fig:bifdiag_general}, in this case for fixed $\bphi_2 = -0.85$ and for the particular domain size $L=16\pi$. The various states are characterized by their norm, Helmholtz free energy density, chemical potentials, and grand potential density. Four types of states can be distinguished: (i) The homogeneous (liquid) state, that is stable at small $\bphi_1$ (solid blue line), which becomes linearly unstable at $\bphi_1=-0.5572$ to periodic perturbations of critical wavenumber $q=1$ (dashed blue line). At the linear stability threshold, a branch of (ii) periodic (crystalline) states bifurcates supercritically. It is at first linearly stable (solid red line), but destabilizes shortly after in a secondary pitchfork bifurcation where two branches of localized states emerge subcritically (green lines). By `localized states', we mean states consisting of a patch of one phase within a background of another phase. A typical example consists of a portion of the crystal coexisting with the liquid state. The branch of periodic localized states continues as an unstable state (dashed red line) before becoming stable again at a further bifurcation where the branches of localized states terminate. (iii) The pair of branches of localized states that emerge at the secondary bifurcation are called `symmetric' as they have a ($x\to-x$)-symmetry. One exhibits an odd- and the other an even-number of peaks. They are connected by ladder branches (not displayed), that connect the intertwining pair of branches and consist of (iv) unstable asymmetric localized states. Together, they form a tilted snakes and ladder structure typical for homoclinic snaking \cite{BuKn2006pre} in a system with a conservation law \cite{TARG2013pre, Knob2016ijam}. As the branches snake back and forth, pairs of peaks are added until the domain is filled. Then the branches terminate in another secondary pitchfork bifurcation on the branch of periodic states. The snaking is best appreciated from the plots of the norm [Fig.~\ref{fig:bifdiag_general}~(a)], the chemical potential [Fig.~\ref{fig:bifdiag_general}~(b)] and the grand potential [Fig.~\ref{fig:bifdiag_general}~(d)]. The manner in which the different branches are connected is less easy to discern in the plot of the Helmholtz free energy [Fig.~\ref{fig:bifdiag_general}~(c)] since the different localized states have very similar energies. However, the inset illustrates that subsequent localized states are connected via swallow-tail structures typical for hysteretic transitions between metastable states. Note that effectively such a transition also occurs from the liquid state to the localized crystalline state (via the periodic state). In this way, identifying the existence of the localized states allows one to see how a supercritical homogeneous-to-periodic bifurcation can still give rise to a first order liquid-to-crystalline phase transitions. For further details of such a bifurcation analysis in the context of phase transition behaviour see the extensive studies for one-component systems in Refs.~\cite{TARG2013pre,TFEK2019njp}.

Figure~\ref{fig:bifdiag_domainSize} illustrates how decreasing (to $ L=8\pi$, top panel) and increasing (to $ L=32\pi$, bottom panel) the domain size affects the bifurcation diagram. First we note that a larger [smaller] domain size brings the secondary bifurcation where the localized states emerge from the periodic one closer to [further away from] the primary bifurcation. In the thermodynamic limit of an infinite domain, the primary and secondary bifurcations coincide \cite{BuKn2007c}.
In a larger [smaller] domain it takes more [fewer] peaks to fill the domain, resulting in more [fewer] folds in the snaking curve and therefore a denser [less dense] snaking region in the bifurcation diagrams. Ref.~\cite{TFEK2019njp} explains in detail how the increase in the number of folds ultimately results in the emergence of the Maxwell construction when approaching the thermodynamic limit. However, here, in contrast to in Ref.~\cite{TFEK2019njp}, the localized states in Figs.~\ref{fig:bifdiag_general} and~\ref{fig:bifdiag_domainSize} do not converge to a Maxwell construction when merely increasing the domain size. This is because the mean concentration $ \bphi_2 $ is held fixed, i.e., no coexisting liquid and crystalline states with \textit{both} chemical potentials and the grand potential equal can emerge. However, this observation does not apply for the localized states on the snaking branches since in this case the concentrations in each of the two phases can achieve their coexisting concentration values by adjusting the spatial size of each of the two phases in the system. To approach the Maxwell constructions from a bifurcation diagram when increasing the domain size for the present two-field system in a way similar to Ref.~\cite{TFEK2019njp}, one would need to fix one chemical potential, instead of a concentration. Physically, this corresponds to a system with a semipermeable membrane that enables exchange with a particle bath of that species.

\section{Determining coexistence by continuation}
\label{sec:calccoex}

\subsection{General considerations}\label{sec:general_considerations}

The two approaches for calculating phase coexistence that we compare share a similar general idea, which is outlined next. As mentioned, both approaches rely on numerical continuation techniques. Firstly, we describe in section~\ref{sec:one-mode} how the well established one-mode approximation is practically used in conjunction with numerical continuation of solutions of an algebraic equation system. Secondly, section~\ref{sec:coexistence-continuation} introduces our fully nonlinear approach that essentially corresponds to the parallel numerical continuation of two solutions to a system of partial differential equations. 

\begin{figure}
\centering
\def\svgwidth{.8\textwidth}
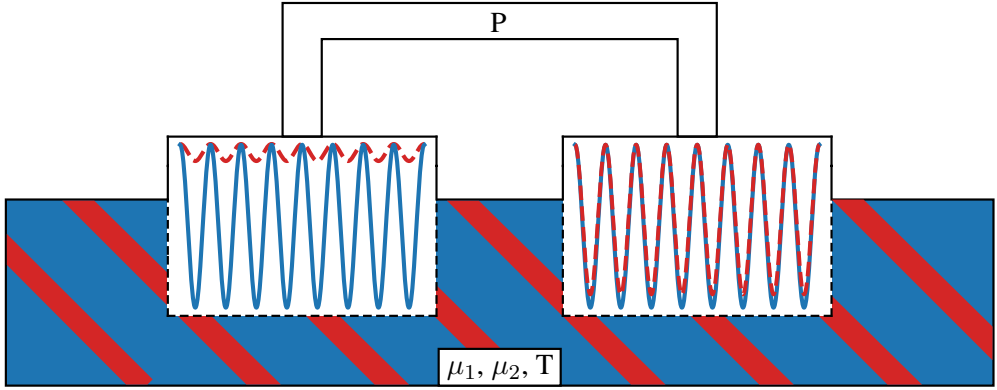
\caption{Sketch illustrating the approach taken by our algorithm: Two simulation boxes in a common bath, that keeps the two boxes at equal pressure, temperature and chemical potentials. }
\label{fig:sketch2boxes}
\end{figure}

In general, coexistence of two distinct phases is characterized by equal temperature, pressure and chemical potentials of all species in the two phases. The system considered here is isothermal, with the parameter $r$ effectively being the temperature. Thus, having set the value of $r$, the temperature is the same in all phases by definition and is therefore not considered further. The pressure $p$ is related to the grand potential density \eqref{eq:gpd} as
\begin{align}
\bar{\omega}= -p,
\end{align}
while the chemical potentials $\mu_j=\delta F / \delta\phi_j$ are introduced in Eqs.~\eqref{eq:twice_integrated}. As a result, the conditions for coexistence of two phases A and B are
\begin{align}
\bar{\omega}^A = \bar{\omega}^B,\label{eq:coex_cond_start}\\
\mu_1^A = \mu_1^B,\\
\textrm{and}\qquad\mu_2^A = \mu_2^B.\label{eq:coex_cond_end}
\end{align} 
From hereon we use superscripts to denote the phase and subscripts to refer to the specific species. Note that Eqs.~\eqref{eq:coex_cond_start}--\eqref{eq:coex_cond_end} correspond to the common tangent construction. In 1d we observe four different phases: (i) the uniform liquid, (ii) a periodic phase where the amplitude of the oscillations in $\phi_1$ are much larger than in $\phi_2$, so we refer to this as the $\phi_1$-crystal, (iii) a corresponding $\phi_2$-crystal, where the oscillations are much larger in the species 2 concentration field and (iv) a crystalline phase where the amplitude of the concentration variations in both fields are comparable, which here we call the ``alloy'' phase. The 2d phases are discussed below in section~\ref{sec:2D_results}.

\begin{figure}[ht!]
\centering
\vspace{-20pt}
\includegraphics[width=\textwidth]{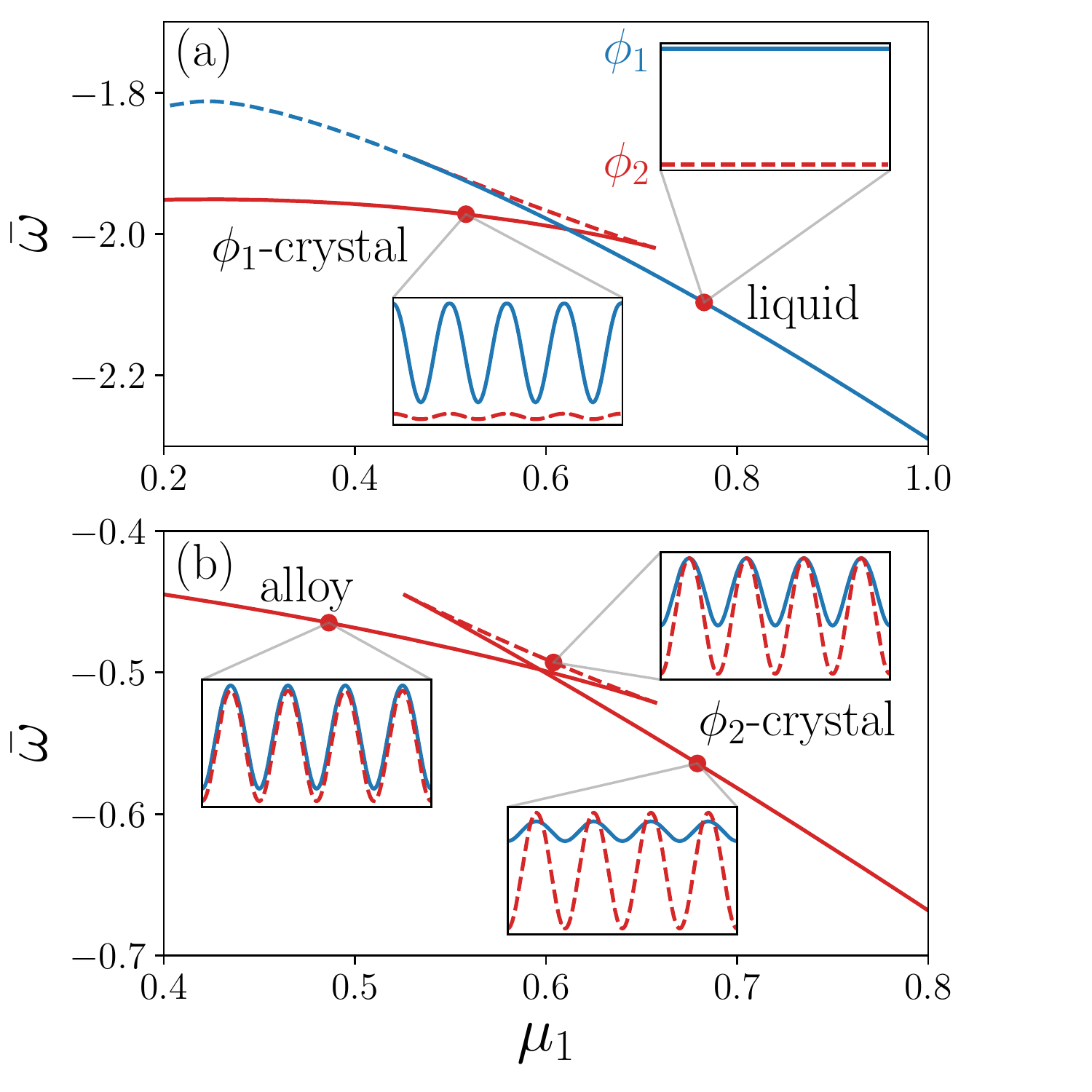}
\caption{Bifurcation diagrams showing branches of (a) the liquid and $ \bphi_1 $-crystal states for $ \mu_2 = -2.0 $ and (b) different crystalline states for $ \mu_2 = 0 $ in 1d, characterized by the grand potential density $\bar\omega$ as a function of the chemical potential $ \mu_1$. The diagram corresponds to a central horizontal crossing of the right half of Fig.~\ref{fig:om-full_binodals_mu1_mu2}~(a) below. In (a), the crystalline state (red lines) emerges subcritically from the liquid branch (blue lines). The dashed portions of the lines show where the states are unstable. The insets show concentration profiles at the positions marked by filled circles. In (b), the two stable states (alloy and $\phi_2$-crystal, red lines) are connected through a swallow-tail structure.  The three insets show concentration profiles  at the positions marked by filled circles. The branches are obtained by continuation of the fully nonlinear model for a single box. A two-parameter continuation of the loci of the two saddle-node bifurcations in (b) is shown in Fig.~\ref{fig:foldconts}. The remaining parameters and line styles are as in Fig.~\ref{fig:bifdiag_general}.}
\label{fig:omega_mu1}
\end{figure}

The coexistence of two phases is calculated by employing two parallel simulation boxes (solution domains), as illustrated in Fig.~\ref{fig:sketch2boxes}. The conditions of equal temperature and chemical potentials for the two boxes are ensured via an ``external bath'', i.e.\ by setting the system parameters $r^A=r^B$ and by equilibrating the chemical potentials by adjusting the values of $\bar\phi_j^A$ and $\bar\phi_j^B$. Normally, the quantities in one box are chosen and the ones in the other box are accordingly adapted to fulfil the above conditions. The pressure is equilibrated via an auxiliary constraint. Before following a line of two-phase coexistence (i.e.\ a binodal), the corresponding coexisting states are initialised by performing single-parameter single-box continuations, as used to obtain bifurcation diagrams. As illustrated in Fig.~\ref{fig:omega_mu1}~(b), for the alloy and $\phi_2$-crystal phases, there is a first order transition between these two states and they typically coexist over a small parameter range of bistability, e.g., in Fig.~\ref{fig:omega_mu1}~(b) for $0.53\gtrsim\mu_1\lesssim0.66$. 

Fig.~\ref{fig:omega_mu1} displays two bifurcation diagrams, showing $\bar{\omega}$ as a function of the control parameter $\mu_1$, for fixed $\mu_2$. In this representation, a point on the binodal can be directly identified as the crossing point of two sub-branches. Once a point on the binodal is identified, we then follow the binodal in a dedicated continuation run while keeping all conditions \eqref{eq:coex_cond_start}--\eqref{eq:coex_cond_end} fulfilled. The results from this are the values along the binodal of $\mu_j$, $\bar{\omega}$, $\bar\phi_j^A$ and $\bar\phi_j^B$. Subsequently, pairs of these can then be displayed in various representations; the most common being in the $(\mu_1,\mu_2)$- and $(\bar\phi_1,\bar\phi_2)$-planes.

\begin{figure}[ht!]
\centering
\includegraphics[width=0.7\textwidth]{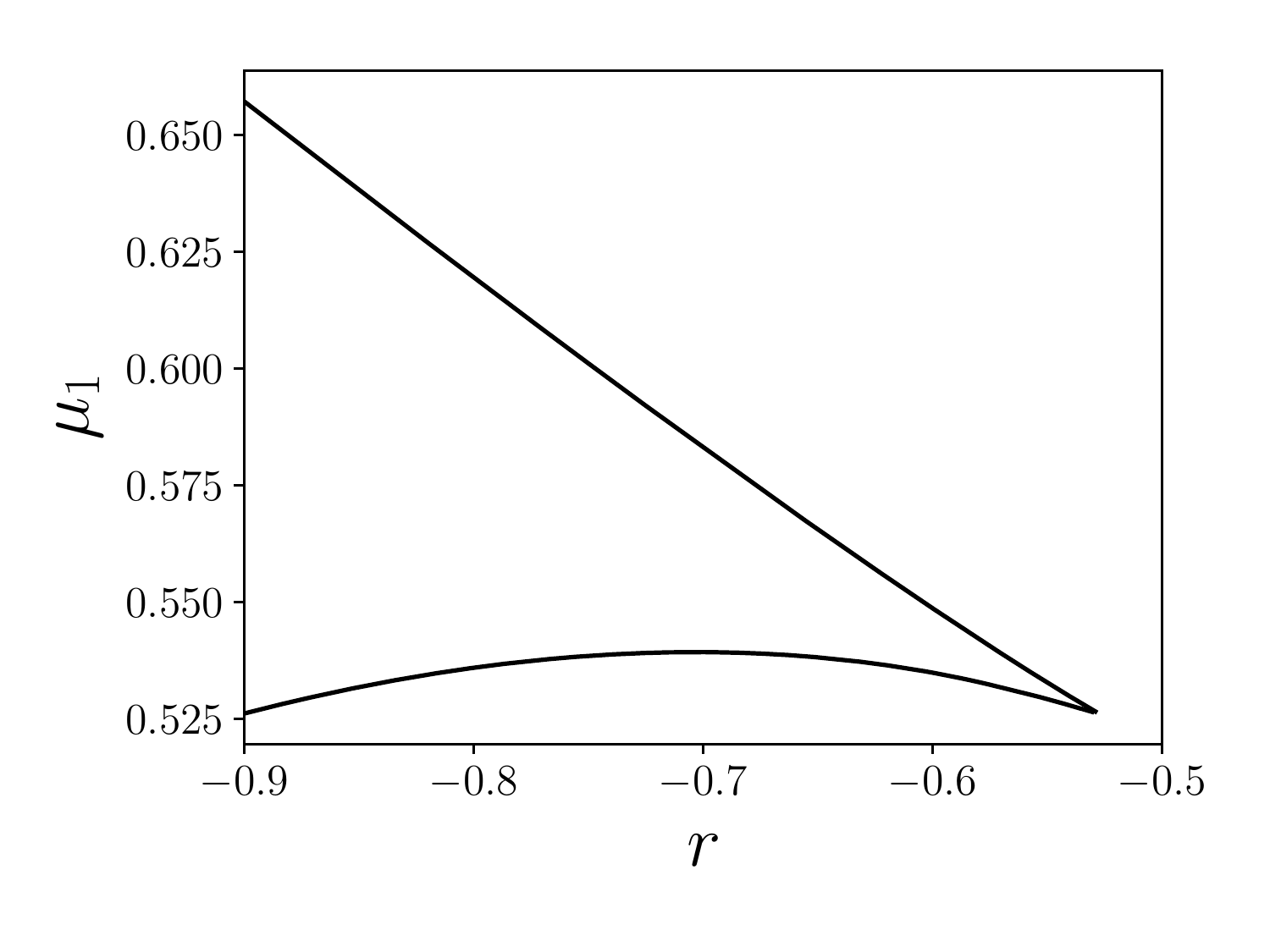}
\caption{Loci of the saddle-node bifurcations of the solution branch in Fig.~\ref{fig:omega_mu1}~(b) are shown in the ($r, \mu_1$)-plane. As the two lines approach with increasing $r$, the swallow-tail structure in Fig.~\ref{fig:omega_mu1}~(b) gets smaller, until the saddle-node bifurcations meet and annihilate in a hysteresis bifurcation. As a consequence, there is no longer a point where distinct alloy and $\phi_2 $-crystal coexist and the phases continuously morph into each other. The remaining parameters are as in Fig.~\ref{fig:omega_mu1}~(b).}
\label{fig:foldconts}
\end{figure}

The procedure works for both the one-mode approximation, where the system of equations to be solved is algebraic, or for the fully nonlinear case, where it consists of ordinary (1d) or partial (2d) differential equations (ODE and PDE, respectively). To identify the binodal starting point, we introduce an auxiliary continuation parameter $ \delta\omega = \omega_B - \omega_A$, which is the difference between the grand potentials of the two phases. With this additional condition, in the initial run we perform a two-parameter continuation in $\delta\omega$ and one of the chemical potentials, and stop when $\delta\omega = 0$, i.e. when we have reached the binodal. Then, in the following run, $\delta\omega=0$ is used as constraint, and a two-parameter continuation in the two chemical potentials directly follows the binodals in parameter space. 
	
Note that the swallow-tail structure in Fig.~\ref{fig:omega_mu1}~(b) is formed of two branches of linearly stable phases and a branch of unstable states connecting them [similar to the structures displayed in Fig.~\ref{fig:bifdiag_general}~(c)]. Starting at the saddle-node bifurcation where the sub-branch of $\phi_2$-crystal ends, it consists of unstable states with monotonically increasing amplitude of the oscillations in the $\phi_1$-field. Similar swallow-tail structures are also visible in the phase diagrams shown below, where they connect binodal lines. A hysteretic transition between two phases of identical symmetry, such as the one between the alloy and the $\phi_2$-crystal in Fig.~\ref{fig:omega_mu1}, corresponds in the thermodynamic limit to a first order phase transition. Following the loci of the saddle-node bifurcations (displayed in Fig.~\ref{fig:omega_mu1}) in a two-parameter continuation, we obtain Fig.~\ref{fig:foldconts}. It shows the loci in the ($r, \mu_1$)-plane. We see that on increasing $r$ for fixed $\mu_2=0$, they approach each other until they annihilate in a hysteresis bifurcation at $r\approx-0.53$, thus terminating the phase coexistence between the $\phi_2$-crystal and the alloy. In the thermodynamic limit, this corresponds to a critical point at this value of $r$, where the first order phase transition ceases to exist. At higher temperatures (higher $r$ values) one can go smoothly from the alloy into the $\phi_2$-crystal, without passing through a distinct phase boundary. This is further discussed below.

In Fig.~\ref{fig:omega_mu1}~(a) we display corresponding branches for the liquid-to-crystal transition, for $ \mu_2 = -2$. The coexistence region is larger as compared to the one in Fig.~\ref{fig:omega_mu1}~(b). In contrast to the transition via two folds in a swallow-tail structure observed for the crystal-to-alloy transition, the liquid-to-crystal transition occurs via a subcritical (symmetry breaking) pitchfork bifurcation where the crystalline state emerges from the liquid one. Similarly, in the equivalent case in the one-component PFC model, at some higher temperature the pitchfork bifurcation becomes supercritical, and the liquid-to-crystal transition changes its character from first order to second order transition. 

For the binodal lines to be calculated in the thermodynamic limit, i.e., for an infinite size domain, the simulation boxes for the crystalline phases have to correspond to an integer number of unit cells of the crystal (one unit cell is sufficient). Thus, the domain size has to be chosen such that it minimizes the Helmholtz free energy at each step. However, since in practice the difference between the results obtained using an approach with a variable domain size and with a fixed one having the size of the unit cell at onset is too small to be resolved on the scale of our figures, we keep the domain size fixed at that value.

\subsection{One-mode approximation}\label{sec:one-mode}

The one-mode approximation is probably the most widely used approach applied to determine phase diagrams of PFC-type models. For the crystalline phases in 1d, the fields are approximated by the harmonic ansatz $ \phi_j \approx \bar{\phi}_j + A_je^{ikx} + c.c. $ with complex amplitudes $ A_j$ and $ c.c.$ denoting the complex conjugate. This ansatz automatically allows for a phase shift between the fields. Introducing it into Eq.~\eqref{eq:cPFC3_FE} and integrating over a unit cell yields the mean free energy:
\begin{align}
\bar{f}  =& \sum_j \left\lbrace\frac{\bar{\phi}_j^2}{2}(r + q^4) + [r + (q^2 - k^2)^2]A_j\bar{A}_j + \frac{1}{4}(6A_j^2\bar{A}_j^2 + 12A_j\bar{A}_j\bar{\phi}_j^2 + \bar{\phi}_j^4)\right\rbrace\notag\\
	&+ c(A_1\bar{A}_2 + \bar{A}_1A_2 + \bar{\phi}_1\bar{\phi}_2).\label{eq:om_FE}
\end{align}
Minimizing Eq.~\eqref{eq:om_FE} with respect to $k$ yields the critical wavelength $k_c = q$. A further minimization with respect to the real and imaginary parts of the amplitudes gives the amplitude equations
\begin{align}
0 &=rA_1 + 3A_1|A_1|^2 + 3A_1\bar{\phi}_1^2 + cA_2\nonumber\\%\label{eq:ampsfirst}\\
0 &=rA_2 + 3A_2|A_2|^2 + 3A_2\bar{\phi}_2^2 + cA_1\label{eq:ampslast}
\end{align}
For the liquid state they are trivially solved ($A_j=0$). The constraints 
\begin{align}
0 = (r + 1)\bphi_1 + 6|A_1|\bphi_1 + \bphi_1^3 + c\bphi_2 - \mu_1\nonumber\\ %\label{eq:om_mu1}\\ 
0 = (r + 1)\bphi_2 + 6|A_1|\bphi_2 + \bphi_2^3 + c\bphi_1 - \mu_2\label{eq:om_mu2}
\end{align}
relate the chemical potentials to the values of $ \bar{\phi}_j $. The grand potential density is obtained using Eq.~\eqref{eq:gpd} with Eqs.~\eqref{eq:om_FE} and \eqref{eq:om_mu2}.

In general, at fixed parameters there can be several solutions to the amplitude equations \eqref{eq:ampslast}. This reflects that in certain parameter regions several phases can simultaneously exist, e.g., the alloy and $ \phi_j $-crystal phase. Of course, this is a condition for the coexistence of phases. The main task is to identify the physically relevant states out of the possibly many solutions of the four coupled algebraic Eqs.~\eqref{eq:ampslast} for the real and imaginary parts of the $A_j$.
Here, we employ numerical continuation as a convenient tool. This allows us to use Eqs.~\eqref{eq:om_mu2} either to measure the $\mu_j$ or to impose them via a constraint. The same applies for the equation defining the grand potential.

Hence, the general continuation procedure described in section~\ref{sec:general_considerations} is specified as follows: Starting from the liquid state, i.e., $A_j = 0$ for particular values of the $\bphi_j$ with corresponding $\mu_j$ and $\omega$, we keep one of the chemical potentials fixed, e.g., $ \mu_2 $, and change the other one, e.g., $\mu_1$, adapting the $\bphi_j$ and $\omega$ in the process. Thus we calculate the branch of liquid states. Eventually, a pitchfork bifurcation is detected where the liquid state changes stability. At this point we switch to the emerging branch, which is a physically relevant periodic (crystalline) state corresponding to a nontrivial solution of the amplitude equations \eqref{eq:ampslast}. Then a binodal point is identified as described in section~\ref{sec:general_considerations}. Subsequently, the binodal-tracking routine is then initialized by populating the two boxes with the two determined coexisting states. Each box has its own set of amplitude equations [Eqs.~\eqref{eq:ampslast}] and constraints [Eqs.~\eqref{eq:om_mu2}, and -- depending on the phases considered -- translation constraints]. To follow the binodal lines, the mean grand potential $ \bar{\omega} $ and both chemical potentials $ \mu_j$ of the two states are held at identical values. Therefore, in total 14 [15, if both states are crystalline] algebraic equations are solved in parallel to determine the 15 [16, if both states are crystalline] unknowns, one of which is used as the control parameter. Typical results are given below in section~\ref{sec:results}.

\subsection{Fully nonlinear approach}\label{sec:coexistence-continuation}
In the fully nonlinear approach we also calculate the phase diagram using numerical continuation techniques. However, in this case it is done by solving coupled ODEs [Eqs.~\eqref{eq:twice_integrated} in 1d] and coupled PDEs [Eqs.~\eqref{eq:twice_integrated} in 2d]. Auxiliary conditions/constraints are employed to break the translational symmetry in systems with periodic boundary conditions. Additional constraints enforce the coexistence conditions of equal pressure [Eq.~\eqref{eq:coex_cond_start}]
\begin{align}
0 &= \bar{\omega}_A - \bar{\omega}_B + \delta\omega, \label{eq:equal_pressure}
\end{align}
and of equal chemical potentials. These determine the mean concentrations. As described above, the auxiliary continuation parameter $ \delta\omega $ allows us to employ continuation to equilibrate the pressure. For the numerical continuation of steady states we use Eqs.~\eqref{eq:twice_integrated}, rather than the steady-state equations before integration. This reduces the computational complexity as fewer derivatives need to be calculated. Additionally, it gives direct access to the chemical potentials. Thus, the mean concentrations $ \bar{\phi}_j $ are not used as control parameters but are just measured in order to observe how they vary according to the given chemical potentials. If Neumann boundary conditions are used, the only additional constraint needed is the equal-pressure condition \eqref{eq:equal_pressure}, since the chemical potentials are directly controlled. The system we need to solve is therefore
\begin{align}
0 &= (r + (q_1^2 + \Delta)^2)\phi_1^A + (\phi_1^A)^3 + c\phi_2^A - \mu_1\label{eq:box1}\\
0 &= (r + (q_2^2 + \Delta)^2)\phi_2^A + (\phi_2^A)^3 + c\phi_1^A - \mu_2\label{eq:box2}\\
0 &= (r + (q_1^2 + \Delta)^2)\phi_1^B + (\phi_1^B)^3 + c\phi_2^B - \mu_1\label{eq:box3}\\
0 &= (r + (q_2^2 + \Delta)^2)\phi_2^B + (\phi_2^B)^3 + c\phi_1^B - \mu_2\label{eq:box4}\\
0 &= \bar{\omega}_A - \bar{\omega}_B + \delta\omega
\end{align}
Thinking of the system as consisting of two boxes, Eqs.~\eqref{eq:box1} and \eqref{eq:box2} describe the first box, and Eqs.~\eqref{eq:box3} and \eqref{eq:box4} the second one. Each box contains one of the states of interest, i.e., the states whose coexistence is investigated. As described previously (section \ref{sec:general_considerations}), in a preliminary run, the pressure in the two boxes is equilibrated, i.e., a continuation is performed to find the point where $ \delta\omega = 0 $. Then, a continuation in the chemical potentials at fixed $ \delta\omega = 0 $ follows the binodal in the parameter plane. In the following section we present typical results from this approach.
\section{Phase behaviour of two-field PFC in 1d}\label{sec:results} 
\subsection{Two-phase coexistence and phase diagrams}\label{sec:binodals}
\begin{figure}[ht!]
\centering
\includegraphics[width=0.9\textwidth]{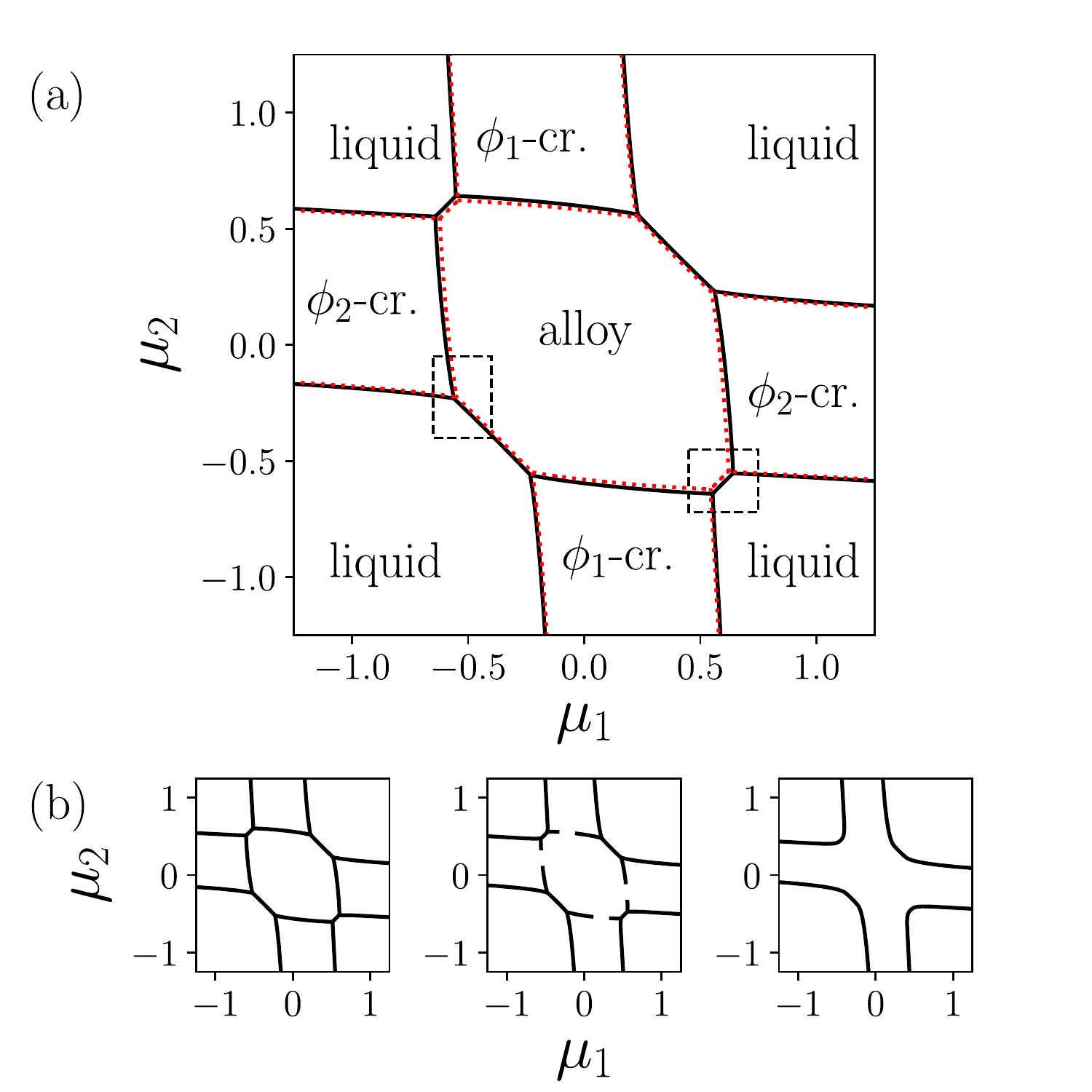}
\caption{Phase diagrams of the binary PFC model in 1d represented in the $(\mu_1$,$\mu_2)$-plane. (a) Phase boundaries calculated using the one-mode approximation (red dotted lines) are compared to the full numerical solution (solid black lines). On the given scale they show good agreement. Regions close to the two triple points bounded by dashed boxes are shown in greater detail in Fig.~\ref{fig:mu1_mu2_zooms}. The parameters are $ r = -0.9 $, $ q_1=q_2=1 $ and $ c = -0.2$. In (b) we display corresponding numerically exact phase diagrams for (from the left) $ r = -0.7 $, $ -0.52 $, and $ -0.3$. The remaining parameters are as in panel~(a). }
\label{fig:om-full_binodals_mu1_mu2}
\end{figure}

\begin{figure}[ht!]
\centering
\includegraphics[width=\textwidth]{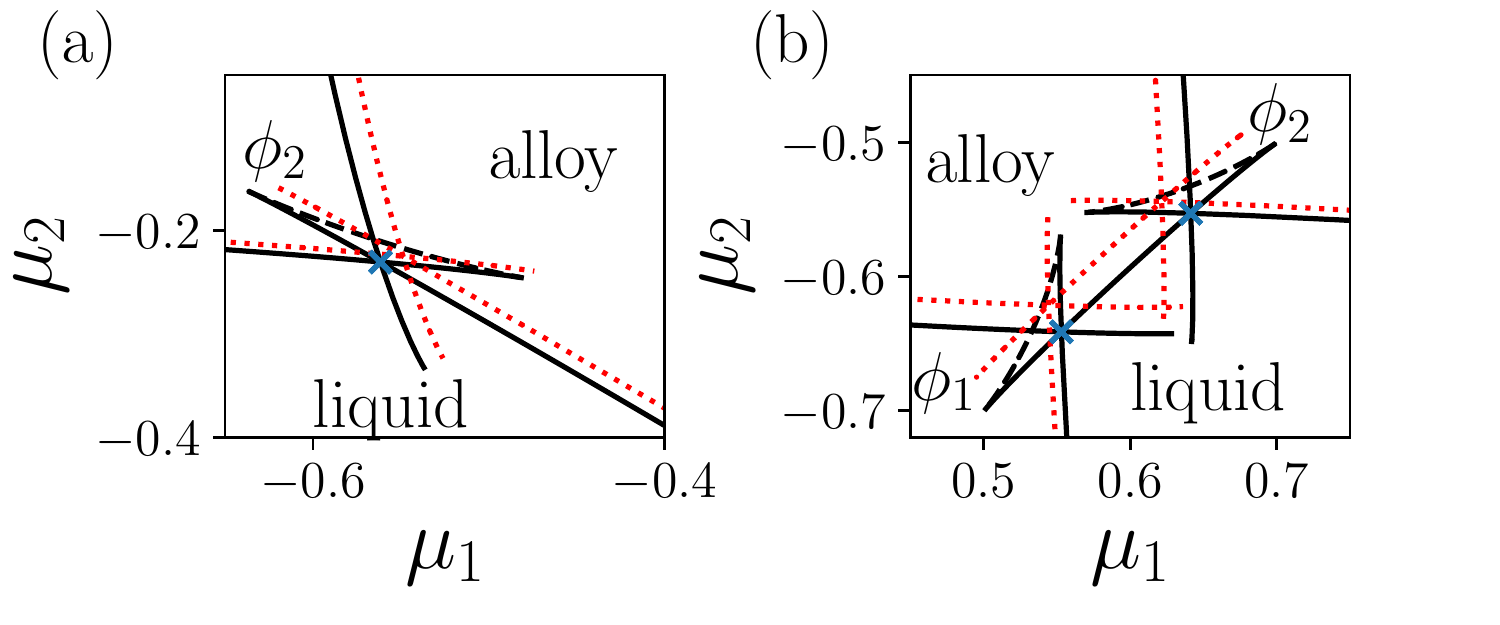}
\caption{Magnifications of regions close to the two triple points marked by dashed boxes in Fig.~\ref{fig:om-full_binodals_mu1_mu2}~(a). The full numerical solution (black lines) and approximate one-mode (red dotted lines) results are shown, as in Fig.~\ref{fig:om-full_binodals_mu1_mu2}~(a), but here enriched with additional details. We see that the that coexistence between two states can extend well into regions where neither of the coexisting states have the lowest free energy. An example is the metastable coexistence of alloy and $\phi_2$-crystal in the area where the liquid is energetically favoured. The triple points are marked by blue crosses (see section~\ref{sec:triple-points} for further detail). Additional dashed lines show ``unstable coexistence'', e.g., in (a) on a line connecting the metastable liquid and $\phi_2$-crystal coexistence with the metastable alloy and liquid coexistence line.}
\label{fig:mu1_mu2_zooms}
\end{figure}

\begin{figure}[ht!]
\centering
\includegraphics[width=\textwidth]{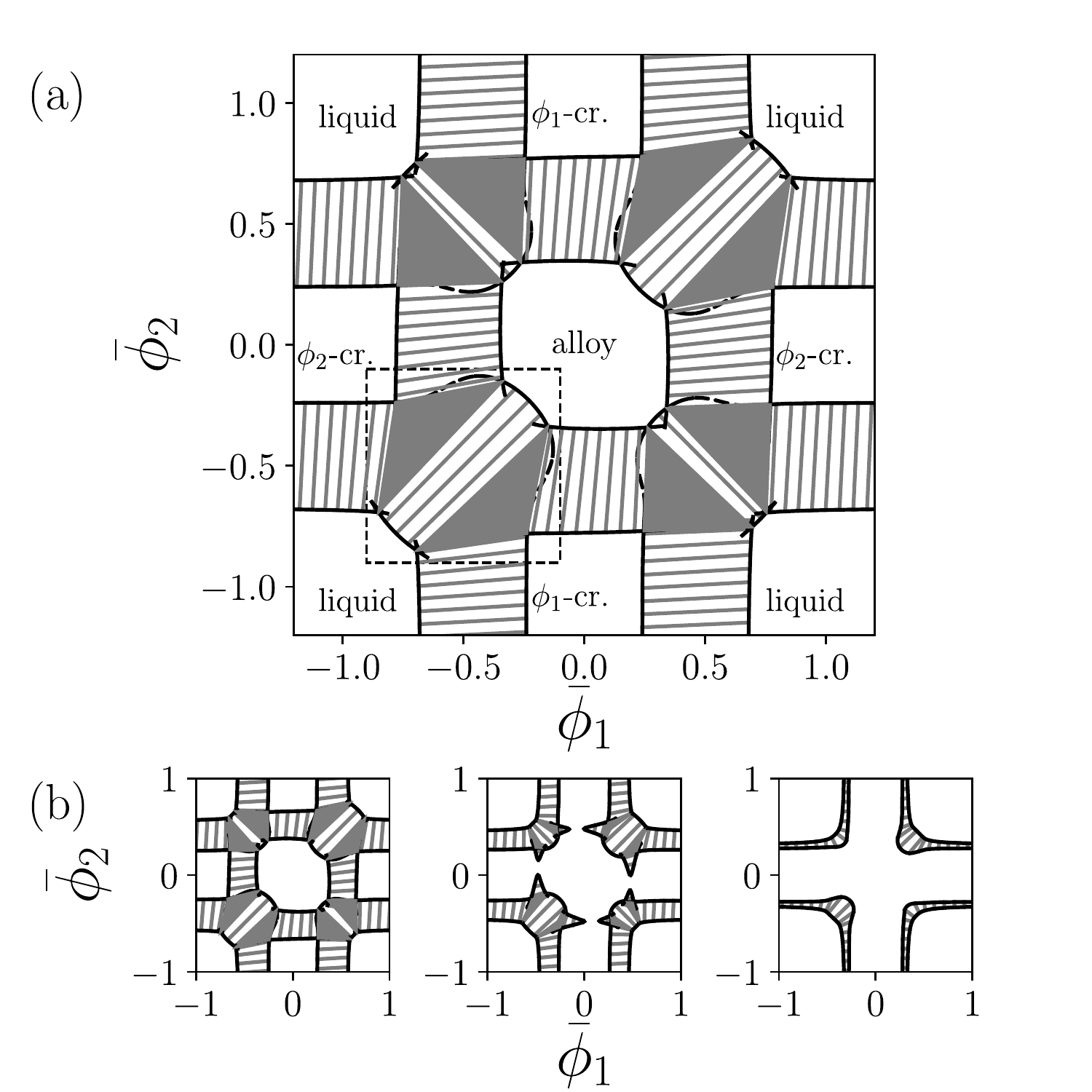}
\caption{Phase diagrams from the full numerical solution of the binary PFC model in 1d displayed in the $(\bar{\phi}_1$,$\bar{\phi}_2)$-plane. The hatched regions correspond to two-phase coexistence of adjacent phases, with the grey tie lines connect coexisting states on the binodals. The triangular grey-shaded areas correspond to three-phase coexistence. Tie lines are only displayed for true thermodynamic coexistence, not for metastable coexistence. The region marked by the dashed box is magnified in Fig.~\ref{fig:phi1_phi2_zooms}. The parameter values are the same as in Fig.~\ref{fig:om-full_binodals_mu1_mu2}. In (b) we display corresponding numerically exact phase diagrams for (from the left) $ r = -0.7 $, $ -0.52 $, and $ -0.3$. The remaining parameters are as in panel~(a).}
\label{fig:phi1_phi2_mulines_r-0.9}
      \end{figure}
      
\begin{figure}[ht!]
\centering
\includegraphics[width=0.8\textwidth]{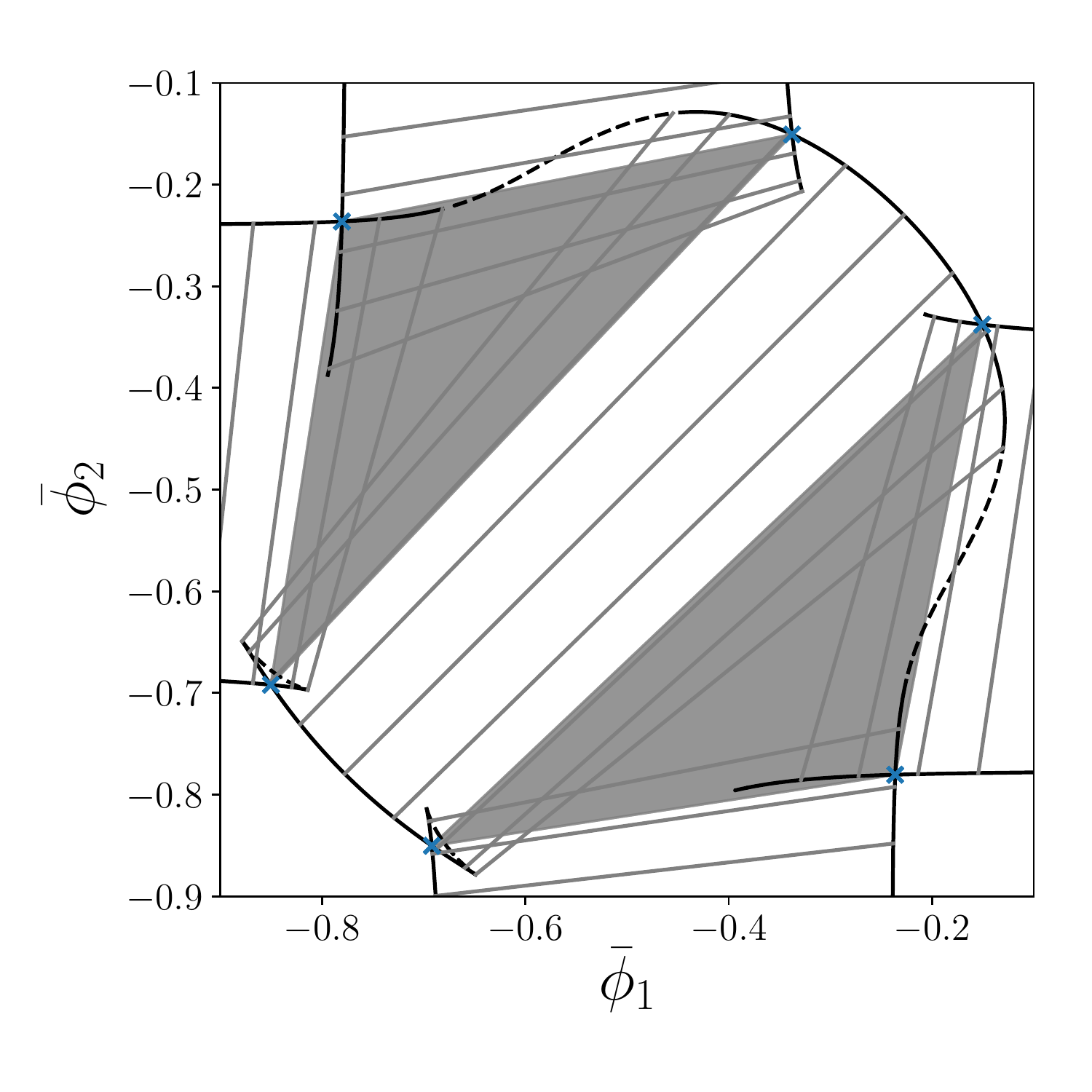}
\caption{Magnification of the area marked by the dashed box in Fig.~\ref{fig:phi1_phi2_mulines_r-0.9}, enriched with further details. In particular, we include several metastable two-phase coexistence that extend beyond triple points and eventually become unstable coexistence (dashed lines). The blue crosses indicate coexisting states of three-phase coexistence -- i.e.\ triple points --see also Sec.~\ref{sec:triple-points}. Note that here, in contrast to Fig.~\ref{fig:phi1_phi2_mulines_r-0.9}, tie lines are also displayed for metastable coexistence.}
        \label{fig:phi1_phi2_zooms}
\end{figure}

\begin{figure}[ht!]
\centering
\includegraphics[width=0.8\textwidth]{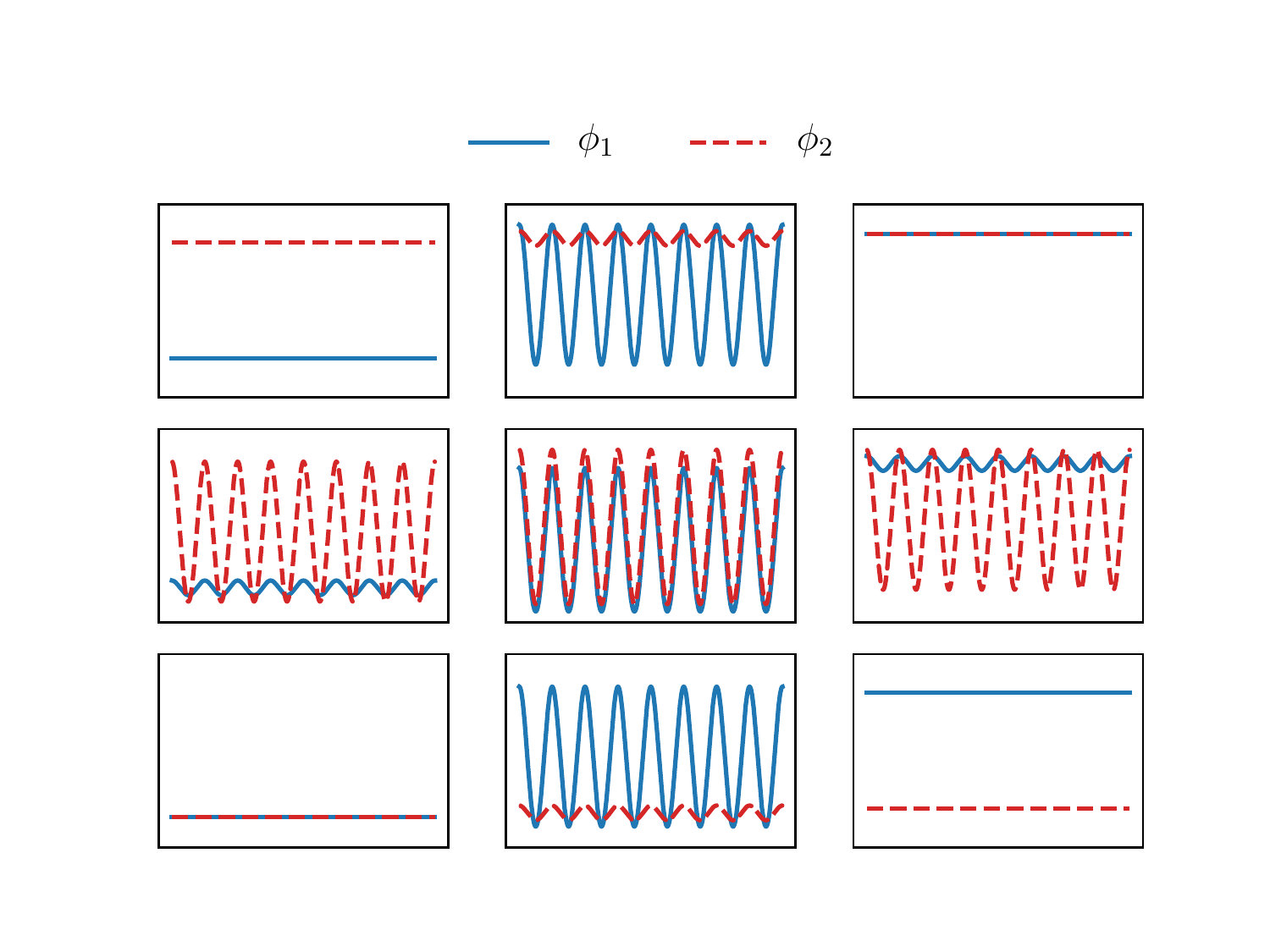}
\caption{Typical concentration profiles for the nine different phases observed in the binary PFC model in 1d. Their positions in the $3\times3$ array corresponds to arrangement of the phases in the phase diagram, displayed in Figs.~\ref{fig:om-full_binodals_mu1_mu2} and \ref{fig:phi1_phi2_mulines_r-0.9}. Namely, the corner profiles represent fluid states, while the central profile is the alloy. The remaining profiles show the four different one-species crystals.}
\label{fig:solution:profiles}
\end{figure}

\begin{figure}[ht!]
\centering
%\maxcomment{Include red lines in Fig.~\ref{fig:om-full_binodals_phibars}?}
\includegraphics[width=0.8\textwidth]{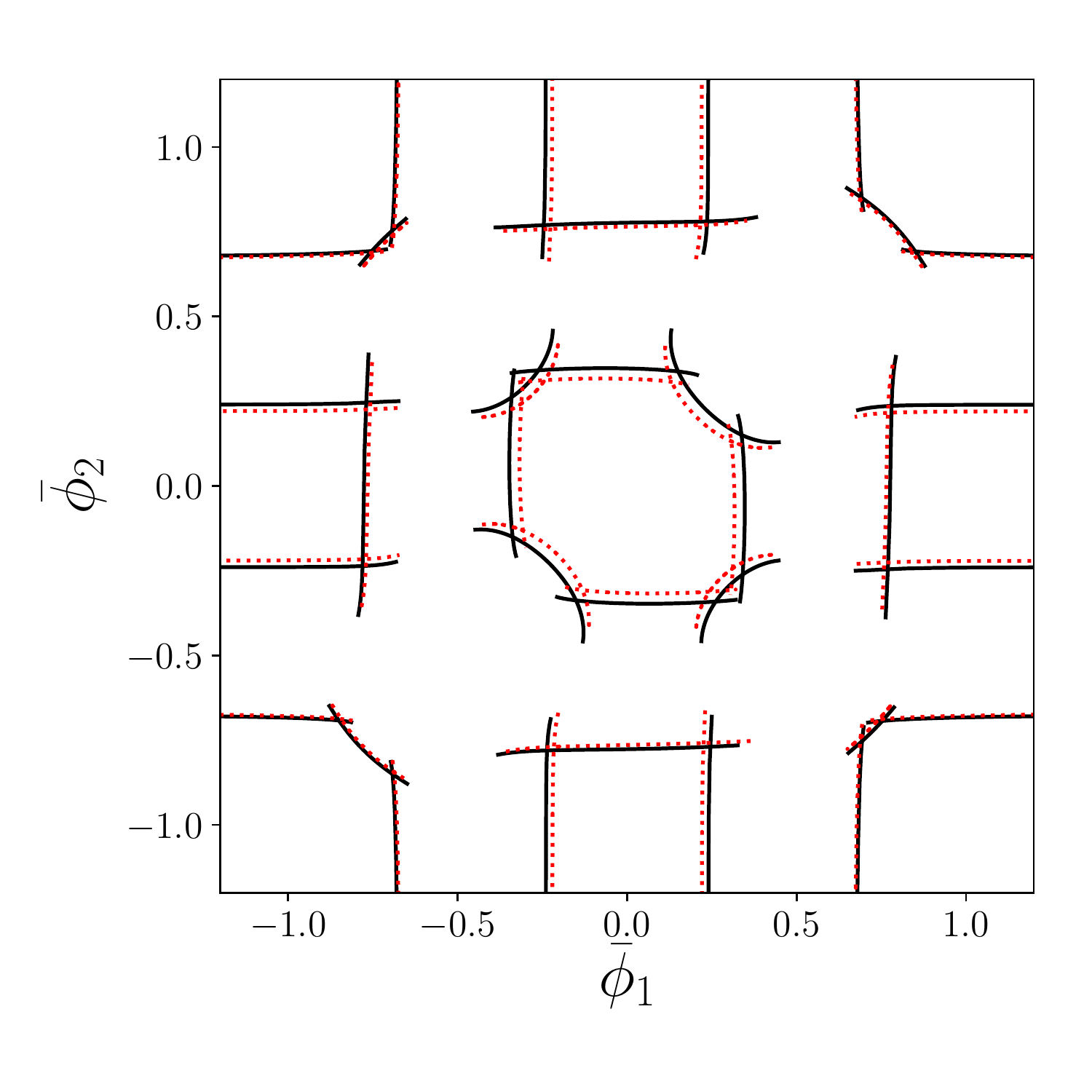}
\caption{Phase diagrams of the two-field PFC model in 1d represented in the $(\bar{\phi}_1$,$\bar{\phi}_2)$-plane. The parameter values are the same as in Fig.~\ref{fig:phi1_phi2_mulines_r-0.9}~(a). Here we compare the phase boundaries calculated using the one-mode approximation (red dotted lines) and the numerically exact ones obtained via our fully nonlinear approach (solid black lines).}
\label{fig:om-full_binodals_phibars}
\end{figure}

Now we present the phase behaviour of the two-field PFC model in 1d thereby comparing the results obtained with the fully nonlinear approach to the one-mode approximation. As effective temperature we first choose $r=-0.9$, as used in the study of the one-field PFC model in Ref.~\cite{TARG2013pre}. Based on the existing results for the one-component model \cite{ElGr2004pre,TARG2013pre}, at such a value we expect a first-order phase transitions to occur between liquid and crystalline states.  Additionally, there are phase transitions between different crystalline states, since we can set the mean concentrations to values where one species forms a crystal, while the other one stays liquid and only shows small modulations due to the coupling. We keep the remaining parameters fixed at $ q_1=q_2=1 $ and $ c = -0.2$.

Figure~\ref{fig:om-full_binodals_mu1_mu2}~(a) show the resulting phase diagram displayed in the $(\mu_1$,$\mu_2)$-plane, while Fig.~\ref{fig:mu1_mu2_zooms} displays magnifications of two particular portions of this phase diagram. Figure~\ref{fig:om-full_binodals_mu1_mu2}~(b) shows how the phase diagram changes as the temperature parameter $r$ is varied. Figure~\ref{fig:phi1_phi2_mulines_r-0.9}~(a) shows the same phase diagram displayed in the $(\bar{\phi}_1$,$\bar{\phi}_2)$-plane, while Fig.~\ref{fig:phi1_phi2_zooms} displays two regions of this in greater detail and Figure~\ref{fig:phi1_phi2_mulines_r-0.9}~(b) shows the variations with $r$. Note that the phase diagrams in both Fig.~\ref{fig:om-full_binodals_mu1_mu2} and \ref{fig:phi1_phi2_mulines_r-0.9} are symmetric with respect to reflections in the diagonals, corresponding to the symmetries of Eqs.~\eqref{eq:cPFC3} discussed above. In Fig.~\ref{fig:solution:profiles} we display a selection of concentration profiles.

In the phase diagram displayed in Fig.~\ref{fig:om-full_binodals_mu1_mu2}~(a), the black solid lines are the coexistence lines obtained via the full numerical solution of the model, while the red dotted lines are the corresponding lines obtained from the one-mode approximation. On the scale of Fig.~\ref{fig:om-full_binodals_mu1_mu2}~(a) the agreement between the two is quite good. However, the magnification in  Fig.~\ref{fig:mu1_mu2_zooms} shows that there are differences in the locations of essential features, which can be up to about 5\% in terms of the chemical potential values. %A similar comparison of the full numerical results and those from the one-mode approximation for the phase diagrams in the $(\bar{\phi}_1$,$\bar{\phi}_2)$-plane can be found in Fig.~\ref{fig:om-full_binodals_phibars}.
In general, the one-mode approximation overestimates [underestimates] the extent of the regions occupied by the liquid [crystalline] phases. As a consequence, the loci of the triple points and related features are also not calculated accurately in the one-mode approximation. For the one-component PFC model, Ref.~\cite{TARG2013pre} shows that a two-mode approximation is sufficient to exactly predict the locus of the tricritical point of the model where the first order transition becomes a second order one. 

In total, there are nine thermodynamically stable liquid and crystal phases. The example concentration profiles in Fig.~\ref{fig:solution:profiles} are arranged in the same way as the phases in the phase diagrams. In the four corners of Fig.~\ref{fig:om-full_binodals_mu1_mu2}~(a) (and Fig.~\ref{fig:solution:profiles}) one finds four fluid phases, namely, (top right) a high-$\phi_1$ {\em and} high-$\phi_2$ liquid phase (short ``high density mixture''); (top left) low-$\phi_1$, high-$\phi_2$ liquid phase (short ``$\phi_2$-liquid''); (bottom left) low-$\phi_1$, low-$\phi_2$ liquid phase (short ``low density mixture''); and (bottom right) high-$\phi_1$, low-$\phi_2$ liquid phase (short ``$\phi_1$-liquid''). In between the four liquid phases there are five crystal phases: (bottom center) and (top center) a $\phi_1$-crystal with low and high, only weakly modulated $\phi_2$ concentration profile, respectively (short ``$\phi_1$-crystal''); (left center) and (right center) $\phi_2$-crystal with low and high, only weakly modulated $\phi_1$ concentration, respectively (short ``$\phi_2$-crystal''); and (center) a crystal of $\phi_1$ and $\phi_2$ peaks that are in phase (short ``alloy''). Note that if we change the sign of the coupling constant so that $c>0$, then the arrangement is anti-phase.

All nine phases can pair-wise coexist with their direct neighbors, giving in total sixteen phase coexistence lines. Some of them are rather short, such as e.g.\ the ones between the alloy on one side and the $\phi_1$-liquid or $\phi_2$-liquid on the other side. The coexistence is best seen in the $(\mu_1$,$\mu_2)$-plane [Fig.~\ref{fig:mu1_mu2_zooms}~(a)] where one can also see clearly how trios of coexistence lines meet at triple points, where all three phases pairwise fulfil the coexistence conditions \eqref{eq:coex_cond_start}-\eqref{eq:coex_cond_end}. Note that all of the triple points consist of the coexistence of a liquid, a one-field crystal and the alloy phase. Leading up to the triple points, the solid lines represent two-phase coexistence in the thermodynamic limit, i.e., the coexistence of two thermodynamically stable phases. However, the continuation method that we use allows us to in addition follow the coexistence beyond the triple points (see the magnifications in Fig.~\ref{fig:mu1_mu2_zooms} for a more detailed look at two examples).

The resulting short pieces of solid line that seem to end slightly beyond the triple points in  Fig.~\ref{fig:om-full_binodals_mu1_mu2}~(a) indicate the coexistence between two states where after the triple point neither are the lowest free energy state. Thus, this portion of the lines correspond to a ``metastable coexistence'' of metastable phases. An example displayed in Fig.~\ref{fig:mu1_mu2_zooms}~(a) is the metastable coexistence of the alloy and $\phi_2$-crystal in the region where the liquid is actually the global energy minimum, i.e., the thermodynamically stable phase. Additional dashed black lines in Fig.~\ref{fig:mu1_mu2_zooms} indicate ``unstable coexistence'', e.g., in (a) on a line connecting metastable liquid and $\phi_2$-crystal coexistence with the metastable alloy and liquid coexistence. On this line, the crystal state corresponds to an unstable state as one field is fully developed while the second field increases in amplitude to reach the alloy state (see Fig.~\ref{fig:omega_mu1}). For clarity the coexistence lines with unstable phases are not shown in Fig.~\ref{fig:om-full_binodals_mu1_mu2} and are only displayed in Fig.~\ref{fig:mu1_mu2_zooms}. In this way, each line of unstable coexistence connects the ends of two lines of metastable coexistence and acts as a threshold (or saddle of the free energy landscape) that has to be crossed when going from a metastable coexistence to a thermodynamically stable one. In particular, when studying the dynamics of phase transitions, e.g., via the motion of fronts, it is important to know which states are metastable and which are linearly unstable. 

Further insight can be gained by inspecting the phase diagram in the $(\bar{\phi}_1$,$\bar{\phi}_2)$-plane, displayed in Fig.~\ref{fig:phi1_phi2_mulines_r-0.9}. In this representation, pairs of coexisting states from along the phase boundaries in Fig.~\ref{fig:om-full_binodals_mu1_mu2} lie on a pair of binodal lines (black solid lines). Particular coexisting states of equal chemical potentials and pressure are connected by tie lines (gray lines), corresponding in each case to a Maxwell construction. In Fig.~\ref{fig:phi1_phi2_mulines_r-0.9} they are only shown for thermodynamically stable coexistence while the magnification in Fig.~\ref{fig:phi1_phi2_zooms} also provides them for the metastable coexistence.
States in the two-phase region between the binodals are unstable w.r.t.\ separation of phases and would evolve along the tie lines. The eight triple points of Fig.~\ref{fig:om-full_binodals_mu1_mu2}~(a) become extended triangular regions (gray shaded areas). States within these regions separate into the three coexisting phases with concentrations $\bar{\phi}_j^l$, $\bar{\phi}_j^c$ and $\bar{\phi}_j^a$ situated at the corners of the respective triangle. Note that this occurs both in the thermodynamic limit and also in finite domains large enough to accommodate regions of all three phases. This results in an intricate bifurcation behaviour to be presented elsewhere. The volume fraction each phase occupies is determined by the choice of parameters $ \bar{\phi}_j $.

The eight triple points merit special attention. They can be determined directly and accurately by extending the two-box method to a three-box method. This then allows one to track the triple points in a space spanned by three parameters as done below in section~\ref{sec:triple-points}.

Figure~\ref{fig:om-full_binodals_mu1_mu2}~(b) shows how the phase behaviour changes with increasing effective temperature, i.e., when increasing $r$ towards less negative values. The three panels display the full numerical solution phase diagrams in the $(\mu_1$,$\mu_2)$-plane for  $ r = -0.7$, $-0.52$, and $-0.3$, respectively. Only the thermodynamically stable coexistence lines are shown. The corresponding sequence of phase diagrams in the $(\bphi_1$,  $\bphi_2)$-plane representation is displayed in Fig.~\ref{fig:phi1_phi2_mulines_r-0.9}~(b).
For $ r = -0.7$ the coexistence lines between the alloy and all four $\phi_j$-crystals still connect the eight triple points and the overall arrangement of the phase diagram is identical to the one at $r = -0.9$ in Fig.~\ref{fig:om-full_binodals_mu1_mu2}~(a). The metastable parts of the coexistence lines have become shorter, i.e., hysteresis becomes less important (not shown). Also the two-phase coexistence regions in the $(\bar{\phi}_1, \bar{\phi}_2)$-plane have decreased in size [see Fig.~\ref{fig:phi1_phi2_mulines_r-0.9}~(b)].  At a certain effective temperature the coexistence lines connecting the triple points break-up at a point on the interval between the triple points. This break-up creates eight critical points at the ends of the lines of first order phase transitions. Increasing $r$, these critical points recede towards the triple points shortening the corresponding lines of coexistence between the alloy and the various other crystals. An example is shown in the panel for $ r = -0.52 $ in Fig.~\ref{fig:om-full_binodals_mu1_mu2}~(b). The finite-size equivalent of such a critical point is the hysteresis bifurcation discussed at Fig.~\ref{fig:foldconts}. 

At the critical points, the mean concentrations $ \bphi_j $ of the two coexisting crystalline states become equal, i.e., the two corresponding binodal lines in Fig.~\ref{fig:phi1_phi2_mulines_r-0.9}~(b) meet. Beyond the critical points, there is no longer a phase transition between the alloy and $ \phi_j $-crystal and one can go smoothly from one to the other.  Further increase of $r$ results in the complete disappearance of the alloy-crystal coexistence lines, where the critical points meet (and annihilate with) the corresponding triple points. An example is the final panel in Fig.~\ref{fig:om-full_binodals_mu1_mu2}~(b). It shows that for $ r = -0.3$ the first order phase transitions between the alloy and $ \phi_j $-crystal have all completely vanished, as have all triple points. The moment when the critical points reached the triple points, these all ceased to exist. As a result, one can go from all the single-species crystals into the alloy state continuously in the whole parameter plane, without encountering any phase transitions. The behaviour of the triple points and three-phase coexistence is discussed further in the next section.

\subsection{Three-phase coexistence and triple-point continuation}\label{sec:triple-points}
\begin{figure}[ht!]
\centering
\includegraphics[width=\textwidth]{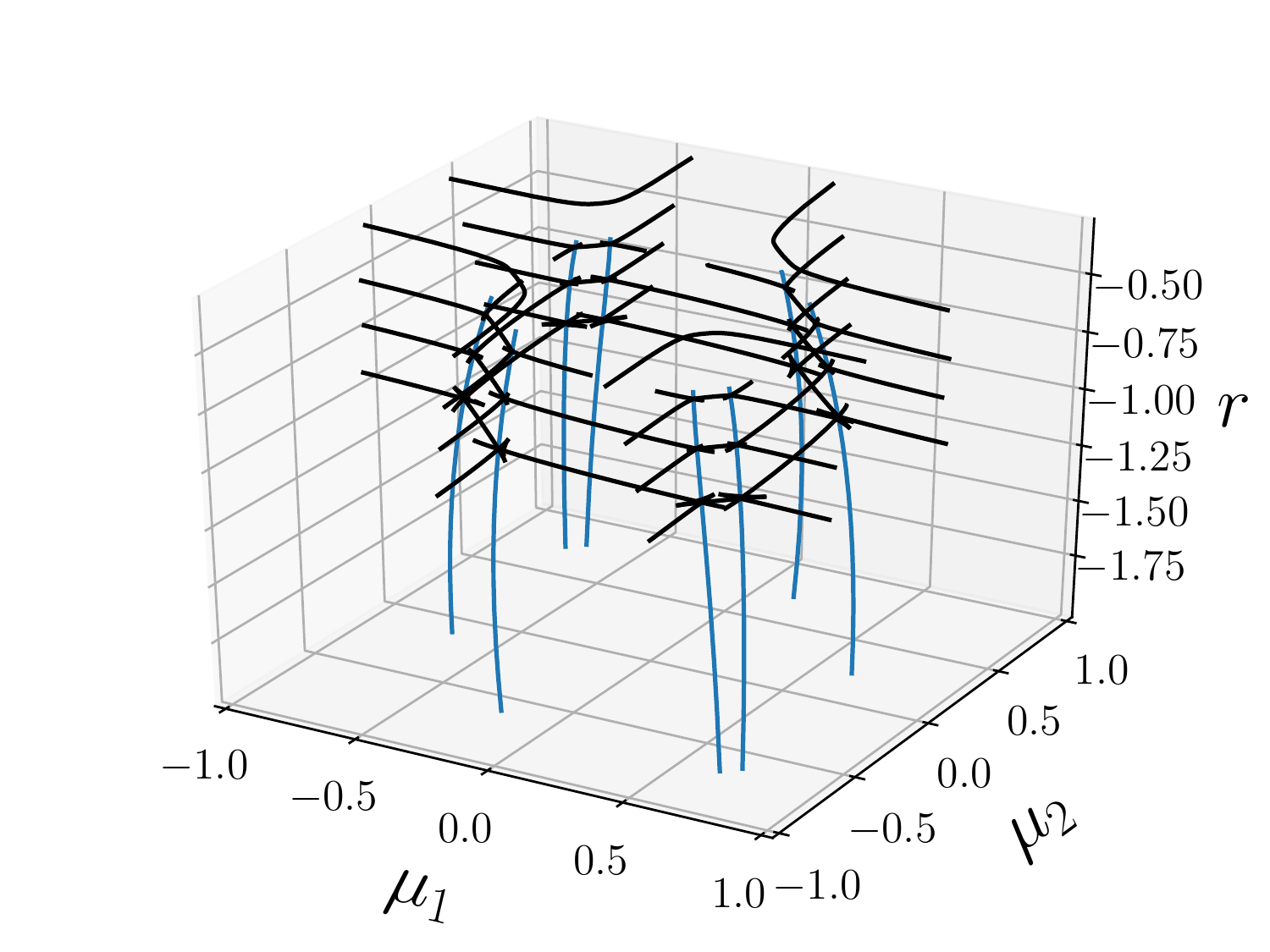}
\caption{Plots of the triple-points (blue lines) and phase boundaries (black lines) in the $(\mu_1, \mu_2)$-plane as the scaled temperature $r$ is varied. The phase boundaries are displayed for $r = -0.9, -0.7, -0.52 $ and $ -0.3 $. The results are all obtained with the fully nonlinear approach. The triple-points cease to exist at higher temperatures (for $r\gtrsim -0.495$ and $r\gtrsim -0.444$), where the transition between the alloy and the various other crystal phases becomes a smooth gradual change and there is no longer a phase transition. The remaining parameters are as in Fig.~\ref{fig:om-full_binodals_mu1_mu2}. }
\label{fig:tripelpoints_mus_r-slices}
\end{figure}

\begin{figure}[ht!]
\centering
\includegraphics[width=\textwidth]{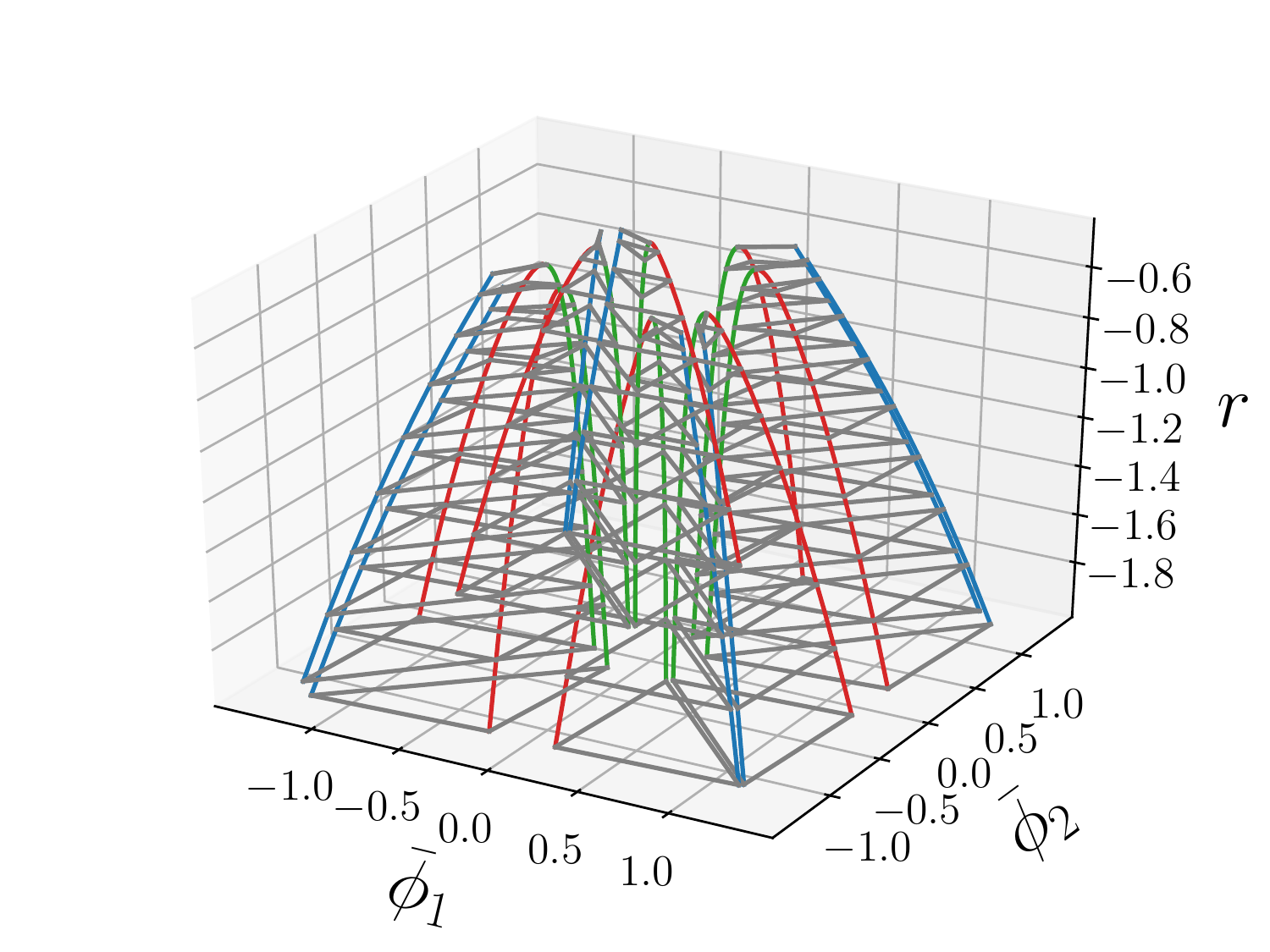}
\caption{The three-phase coexistence regions in the ($\bar{\phi}_1, \bar{\phi}_2, r $)-space (the grey triangles are slices at a sequence of fixed values of $r$), corresponding to the triple-points in Fig.~\ref{fig:tripelpoints_mus_r-slices}. The blue lines correspond to $\bar{\phi}_j^l $, the mean order parameters of the liquid phase at coexistence; the red lines to $ \bar{\phi}_j^c $, the value for the one-species crystals; and the green lines to $ \bar{\phi}_j^a $, the values for the alloy. In Fig.~\ref{fig:tripelpoints_phis_r-slices_3rdquad} we display a magnification of a portion of this diagram. The remaining parameters are as in Fig.~\ref{fig:om-full_binodals_mu1_mu2}.}
\label{fig:tripelpoints_phis_r-slices}
\end{figure}

\begin{figure}[ht!]
\centering
\includegraphics[width=\textwidth]{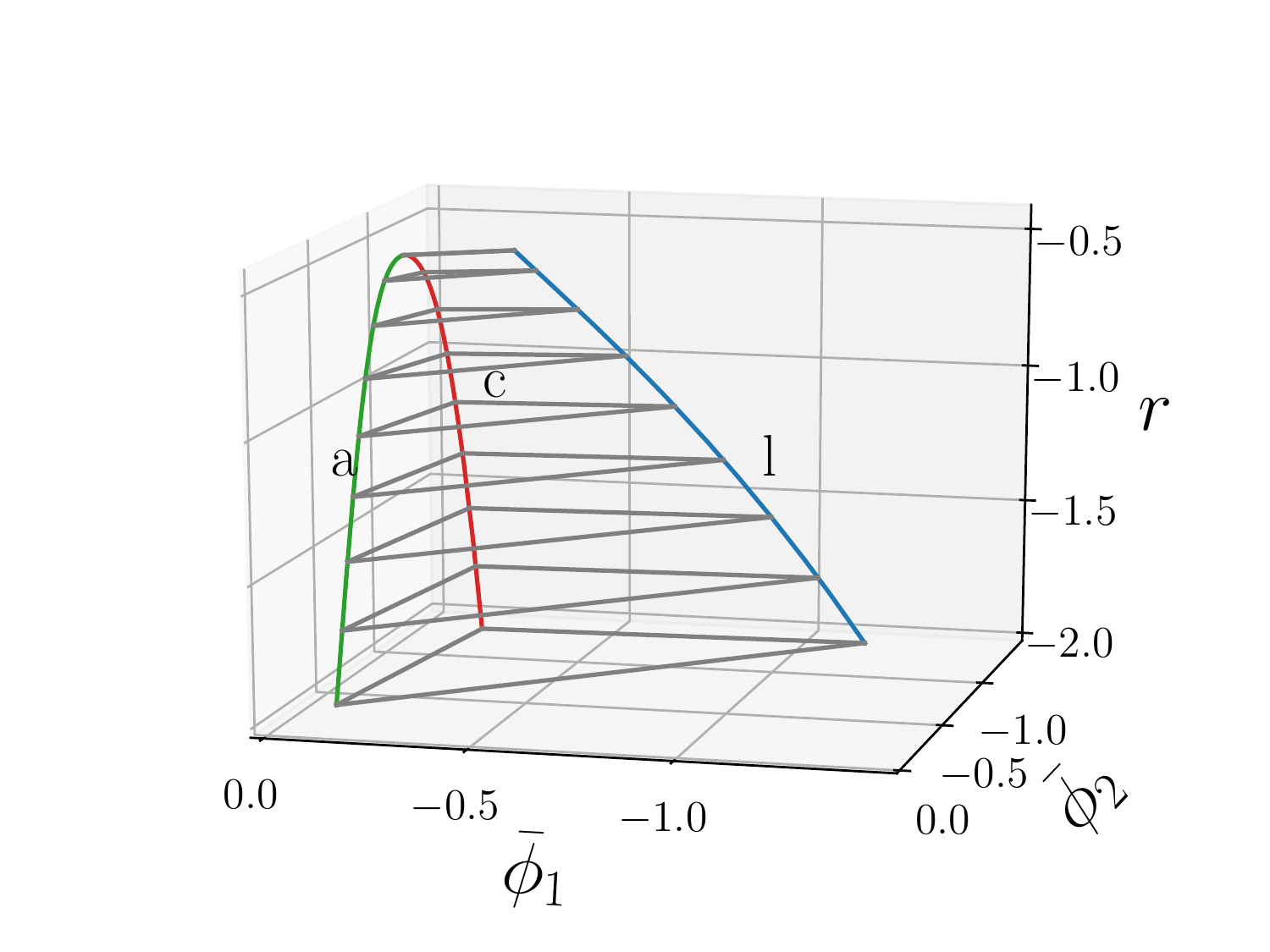}
\caption{A magnification of one of the three-phase regions in Fig.~\ref{fig:tripelpoints_phis_r-slices}. Where the triple-point vanishes at high $r$, the triangular three-phase region reduces to a single line. There, the mean order parameters $ \bar{\phi}_j^c $ and $ \bar{\phi}_j^a $ become equal, as the two coexisting states become identical. Parameters and line styles are as in Fig.~\ref{fig:tripelpoints_phis_r-slices}.}
\label{fig:tripelpoints_phis_r-slices_3rdquad}
\end{figure}

 To track triple points, we extend the two-box continuation method for the determination of lines of coexistence to a three-box method. The fully nonlinear continuation approach then allows us to directly follow them and the associated three-phase regions in an extended parameter spaces spanned either by $ \mu_1 $, $ \mu_2 $ and $r$ or by $ \bar{\phi}_1 $, $ \bar{\phi}_2 $ and $r$. At three-phase coexistence, where two two-phase coexistence lines cross, all three phases have the same pressure and chemical potentials. In the case of the phase boundaries at fixed temperatures studied above, the two-parameter continuation in the chemical potentials $ \mu_1 $ and $ \mu_2 $ is facilitated by the equal pressure constraint. Including a third phase, a second equal pressure constraint is added, to enable us to perform the three-parameter continuation of the triple points.

 Typical results are displayed in Figs.~\ref{fig:tripelpoints_mus_r-slices} and \ref{fig:tripelpoints_phis_r-slices}. They show the loci of the triple points in the ($\mu_1,\mu_2,r$)-space and in the ($\bar{\phi}_1, \bar{\phi}_2, r $)-space, respectively. Figure~\ref{fig:tripelpoints_phis_r-slices_3rdquad} shows a magnification of the three-phase region associated with one of the triple-points.  In addition to the triple points, Fig.~\ref{fig:tripelpoints_mus_r-slices} also displays the two-phase coexistence lines at four fixed $r$-values, identical to the ones used in Fig.~\ref{fig:om-full_binodals_mu1_mu2}.  With increasing temperature the swallow-tail structure becomes smaller until at the temperatures $r\approx -0.495$ and $r\approx -0.444$ the lines of triple points end. This is also clearly visible in the ($\bar{\phi}_1, \bar{\phi}_2, r $)-representation in Fig.~\ref{fig:tripelpoints_phis_r-slices} and the magnification for one of the triple points in Fig.~\ref{fig:tripelpoints_phis_r-slices_3rdquad}.

The coexistence line between alloy and single-species crystal breaks at slightly lower $r$ than the $r$-value where the line of triple points ends. The pairs of critical points created by the break in the coexistence lines then recede towards the triple points at the other ends of the coexistence lines.  At the transition temperature, the triangular three-phase region reduces to a single line because the mean order parameters $ \bar{\phi}_j^a $ for the alloy and $ \bar{\phi}_j^c $ for the single-species crystals become equal, so that the alloy and the crystal become indistinguishable. This reflects the fact that the former triple point becomes just a normal point on a two-phase coexistence line. In other words, this is the point where the pairs of a green and a red line (each pair forming a parabola) in Figs.~\ref{fig:tripelpoints_phis_r-slices} and~\ref{fig:tripelpoints_phis_r-slices_3rdquad} have their apex.
\section{Phase behaviour of the binary PFC in 2d}\label{sec:2D_results}

\begin{table}[ht!]
\caption{
 Nomenclature of phases observed for the two-field PFC model in 2d. Exchanging fields $\phi_1$ and $\phi_2$ gives four additional phases (dl, ul, sl, du) not included here. The abbreviations introduced here are employed in Figs.~\ref{fig:phase_diagram_2D_mus} and \ref{fig:phase_diagram_2D_phis}.}
\label{tab:phase_combination}
\begin{tabular}{l l|c}
$ \phi_1 $& $ \phi_2 $ & \phantom{x}abbreviation \\
\hline
liquid & liquid & ll\\
weakly modulated \phantom{xxx}& down-hexagons \phantom{xxx}& ld\\
weakly modulated & up-hexagons & lu\\
weakly modulated & stripes & ls\\
down-hexagons & down-hexagons & dd\\
up-hexagons & up-hexagons& uu\\
up-hexagons & down-hexagons & ud\\
stripes & stripes & ss
\end{tabular} 
\end{table}

\begin{figure}[ht!]
\centering
\includegraphics[width=0.8\textwidth]{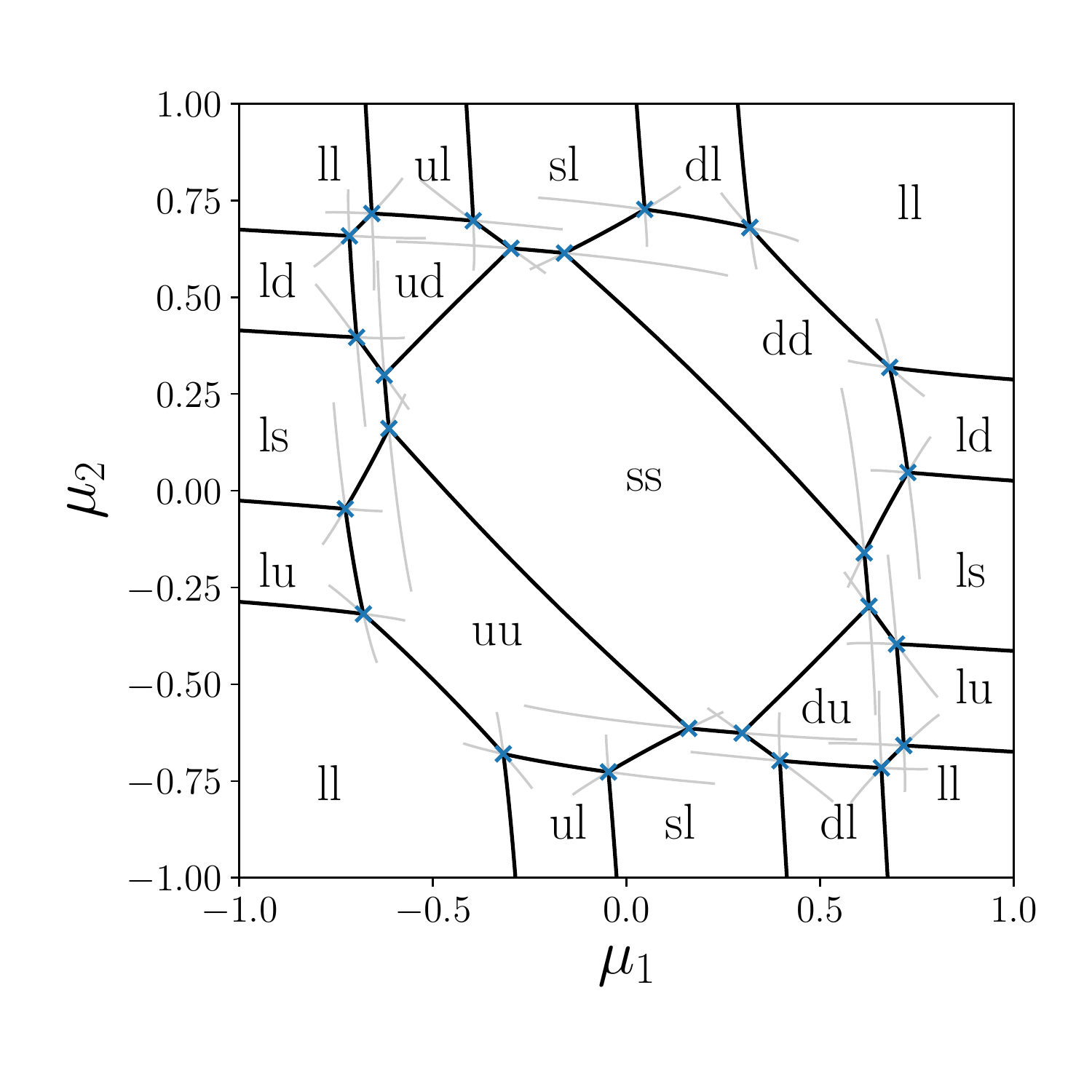}
\caption{The phase diagram of the binary PFC model in 2d in the $(\mu_1$,$\mu_2)$-plane, calculated from full numerical solutions of the model in conjunction with continuation methods. The black lines are the two-phase coexistence lines and the grey lines are their continuations indicating metastable coexistences. The various different phases are indicated using the two-letter abbreviation convention given in Table \ref{tab:phase_combination}. The blue crosses mark the locations of the triple-points, which are calculated with the three-box approach. The remaining parameters are $ r = -0.9 $, $ q_1=q_2=1 $ and $ c = -0.2 $.} \label{fig:phase_diagram_2D_mus}
\end{figure}

\begin{figure}[ht!]
\centering
\includegraphics[width=0.8\textwidth]{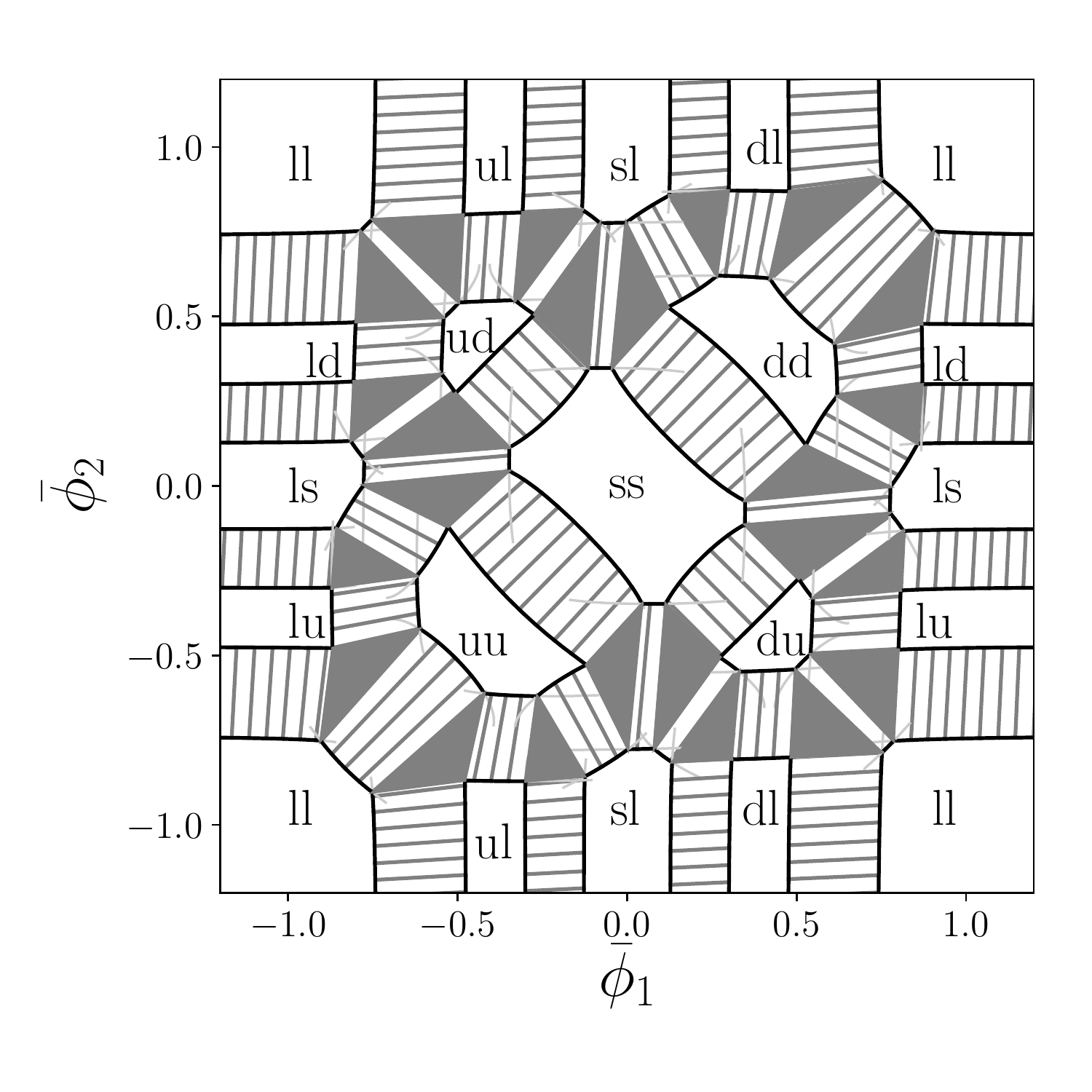}
\caption{The phase diagram of the binary PFC model in 2d displayed in the ($\bphi_1,\bphi_2$)-plane; see also the corresponding diagram in Fig.~\ref{fig:phase_diagram_2D_mus}. The solid black and thin light grey lines denote thermodynamically stable and metastable binodal lines, respectively. Thermodynamically stable coexisting states are connected by dark grey tie lines. Triangular three-phase regions are shaded dark-grey. The phases are abbreviated as defined in Table~\ref{tab:phase_combination}. The parameters are the same as in Fig.~\ref{fig:phase_diagram_2D_mus}.} \label{fig:phase_diagram_2D_phis}
\end{figure}

\begin{figure}[ht!]
	\centering
	\includegraphics[width=0.8\textwidth]{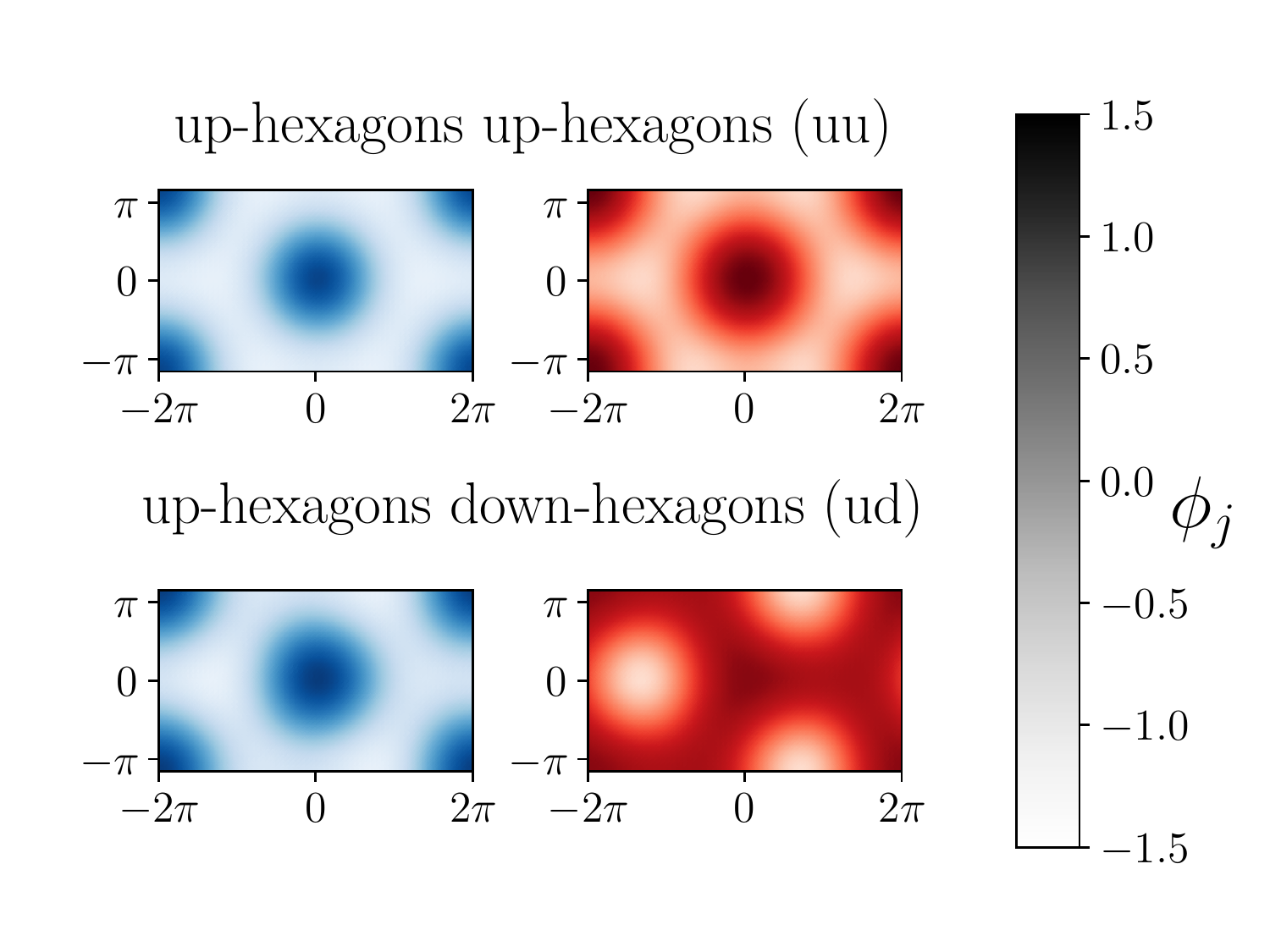}
	\caption{Examples of profiles for the phases that have no 1d equivalent. The left [right] panels give $\phi_1 $ [$ \phi_2 $] in blue [red]. The top panel illustrates the uu-phase where both fields form up-hexagons. In contrast, the bottom panels show the ud-phase where $ \phi_1$ forms up-hexagons and $ \phi_2 $ down-hexagons. Again, exchanging fields $ \phi_1 $ and $ \phi_2 $ gives additional solutions, as does inverting them. The grey scale of the colour bar indicates the concentration values for both fields. } \label{fig:solutions_2D}
\end{figure}

Having described the phase behaviour of the binary PFC model in 1d, we now present results for the phase behaviour of the model in 2d, as obtained via our continuation method and full numerical solutions. For the effective temperature, we again choose $r=-0.9$, the same as used for many of the 1d results presented above. At such a value we expect first-order phase transitions to occur between liquid phases, stripe phases and hexagonal crystalline phases, similar to what is observed in the one-component PFC model \cite{ElGr2004pre,TARG2013pre}. Further, we expect phase transitions between different crystalline hexagonal states, as observed in 1d and additional transitions between different stripe phases. The remaining parameters we keep as before, fixed at $ q_1=q_2=1 $ and $ c = -0.2$. 

Figs.~\ref{fig:phase_diagram_2D_mus} and \ref{fig:phase_diagram_2D_phis} show the resulting phase diagram in the $(\mu_1$,$\mu_2)$- and the $(\bar{\phi}_1$,$\bar{\phi}_2)$-plane, respectively. As in 1d, in 2d both are symmetric with respect to reflections in the diagonals, as expected. We distinguish and denote the various phases based on the patterns that each of the two fields can form: uniform or weakly modulated (liquid states, as in 1d), stripe pattern (as 1d crystal) and hexagonal pattern (cf.~the discussions for one-component models, e.g., in Refs.~\cite{AWTK2014pre,TFEK2019njp}). For the latter there exist two different versions: up-hexagons and down-hexagons with density peaks and density troughs forming the triangular point lattice, respectively.

In the binary PFC model, the various phases known from the one-component system can now appear for $\phi_1$ and $\phi_2$. In principle, any combination is possible, including ``alloys'' of stripes and ``alloys''  of the various hexagons. The thermodynamic phases that actually occur for the parameter values chosen are listed in Table~\ref{tab:phase_combination}, where we do not differentiate between the low and high density liquid (as in 1d). Of particular note is the case where the two uncoupled fields would form up-hexagons and down-hexagons respectively, i.e., the ``ud'' and ``du'' phases. There, the usual in-phase structure is no longer energetically favourable and the two hexagonal patterns are phase shifted and the unit cell is no longer reflection symmetric. Fig.~\ref{fig:solutions_2D} displays examples of concentration profiles for the phases that have no 1d equivalent. Profiles that are not shown can be obtained by applying the various symmetries of the system. 
The abbreviations introduced in Table~\ref{tab:phase_combination} are employed in Figs.~\ref{fig:phase_diagram_2D_mus} and \ref{fig:phase_diagram_2D_phis}.\footnote{Our naming convention is based on notions used in the pattern formation literature, e.g., when discussing patterns for the Swift-Hohenberg equation  \cite{CrHo1993rmp,Hoyle2006}.} Note that the above symmetries imply that under the two reflections, symbols exchange their place and ``u'' becomes ``d'' and vice versa. Note that states other than those present in the phase diagram exist, but they are only metastable or linearly unstable. Nevertheless, they can be studied with the continuation methods we employ. An example is stripes in $\phi_1$ and $\phi_2$ that are orthogonal to each other. They exist as a metastable state in a small central region of the phase diagram. 

Compared to the 1d case where nine phases exist in the phase diagram, the 2d system exhibits many more, namely, twenty-one phases. At $r=-0.9$ all neighbouring phases are separated by first order phase transitions, i.e., in Fig.~\ref{fig:phase_diagram_2D_mus} there are forty-four coexistence lines that form an intricate polygonal network. The twenty-four nodes of this network correspond to triple points of three-phase coexistence. This implies that the accompanying Fig.~\ref{fig:phase_diagram_2D_phis} features forty-four two-phase coexistence regions bounded by pairs of binodals, which in the figure are connected by dark grey tie lines. Furthermore, there are twenty-four grey-shaded triangular three-phase coexistence regions. Overall, a very rich phase behaviour that promises an equally rich selection of possible phase change dynamics, e.g., through moving fronts that replace one state by another neighbouring one.

Note, that at large values of one of $|\mu_j|$ (and correspondingly of one of $|\bar\phi_j|$) the corresponding field is only weakly modulated and effectively decoupled from the other field. Then, a cut through the phase diagram, where the large value is fixed, is equivalent to the phase diagram known for the one-component PFC model. For instance, for increasing chemical potential of the other field, one finds first the low-concentration liquid, then the up-hexagon, stripes, down-hexagon, and finally the high-concentration liquid phase (cf.~Fig.~10 in \cite{TARG2013pre}). 

As outlined above in section~\ref{sec:triple-points}, the locations of the triple points, which are marked by blue crosses in Fig.~\ref{fig:phase_diagram_2D_mus} can be calculated employing the three-box continuation scheme. Again, their loci can be followed in three-parameter space (not shown). 

\section{Concluding remarks}\label{sec:conclusions}

We have presented a method to calculate the phase behaviour of thermodynamic systems using numerical continuation that incorporates obtaining the full numerical solution for the structure of each phase. It can be applied to any continuum model and as an example here has been applied to a binary (two-field) phase-field-crystal model. The model is related to the two-field model introduced in Ref.~\cite{RATK2012pre} where two ``vacancy phase-field-crystal'' models \cite{ChGD2009pre} were coupled. Here we have employed the same coupling. Analogous DDFTs for binary systems and other such PFC models can be found in Refs.~ \cite{ELWG2012ap, AWTK2014pre, SSAH2018jpm, scacchi2020quasicrystal}.

After briefly introducing the binary PFC model and continuation methods in general, we have first employed continuation to determine a number of bifurcation diagrams for finite-size systems and have discussed how the notions from pattern formation and dynamical systems relate to the thermodynamic consideration of phase transitions. This part of our discussion relates to aspects of Ref.~\cite{TFEK2019njp} in the context of two-field models. Second, we have introduced a two-box continuation method that allows one to directly follow lines of phase coexistence, i.e., to follow the Maxwell construction, in the thermodynamic limit, through parameter space. Extending the method to include three-boxes allows one to directly follow triple points in extended parameter spaces. The methods that we have introduced work for both the commonly used simple one-mode approximation, where solutions of an algebraic equation system are followed, and also the full nonlinear model where solutions of a system of ODEs (one-dimensional domain) or of PDEs (two-dimensional domain) are followed.

After introducing the method we have applied it to the binary PFC model, employing on the one hand the one-mode approximation and on the other hand the full nonlinear model. The two approaches were compared in the case of a one-dimensional system. We have shown that the phase diagram obtained via the one-mode approximation agrees qualitatively with that from the full numerical solution, but quantitatively differs, which can be particularly important close to special features of phase diagrams like critical points and triple points. The binary PFC model shows a very rich phase behaviour, e.g., in two dimensions it has twenty-one phases, forty-four lines of two-phase coexistence and twenty-four triple points that can all be obtained and continued quite smoothly with the method that we have developed. Classical methods like direct simulations in time and Picard iteration techniques are cumbersome in comparison and may sometimes end up in metastable states.

Here, we have focused on a binary PFC model in the case of identical particle sizes [i.e., $q_1=q_2=q$ in Eq.~\eqref{eq:cPFC3}]. An even richer phase behaviour is expected when the particle sizes differ -- see e.g.\ the DDFT models in Refs.~\cite{AWTK2014pre, SSAH2018jpm, scacchi2020quasicrystal} and with PFC models in \cite{RATK2012pre}. Such studies may now be supplemented by the systematic determination of phase diagrams employing the techniques presented here. Note that for such systems the domain size also needs to be varied to minimize the Helmholtz free energy. Some of these or other amendments of the model that one can consider often break the model symmetries discussed in section~\ref{sec:binary-pfc} and result in phase diagrams that are not symmetric with respect to the diagonals as observed here. For example, Ref.~\cite{TDME2019prm} presents results for a binary PFC model with up to cubic coupling terms and general quartic single-species energies. Our technique would allow to complete their phase diagrams (Fig.~3 of \cite{TDME2019prm}).

To facilitate the wider application of the proposed method, for the present binary PFC model we have deposited data for most of the calculated coexistence lines and the continuation of the triple points in a data repository \cite{HoAT2020zenodo}. There, also the pde2path routines and suitable starting data for all continuation runs can be found. This shall allow anyone interested to reproduce our results. It should then be straightforward to adapt the codes to study related PFC models, e.g., as used in \cite{RATK2012pre,TDME2019prm}. For the analysis of the present model, worthwhile next steps could be the investigation of cases with $q_1\neq q_2$ and/or different critical points for the two sub-models, i.e., $\Delta r\neq 0$.

The general approach described here can be used to study a broad range of thermodynamic systems. Foremost, these are systems described by continuum models with time evolutions described by a gradient dynamics that can not be solved by exact analytical methods. These include Swift-Hohenberg-like PDE models like the ones used for membranes \cite{WoKA2015pre}, for diblock-copolymer layers \cite{Glas2010sjam, WeKZ2013jcp} and for patterns in Bose-Einstein condensation \cite{HeBD2019pra}. An extensions to integro-differential equations would cover classical DFT models such as those in Refs.~\cite{AWTK2014pre, SSAH2018jpm, scacchi2020quasicrystal}, or even more sophisticated functionals such as the fundamental measure-based DFTs \cite{roth2010fundamental}.

Also, as well as being applicable to all the various PFC models and DDFT models, more generally the approach may be applied to any continuum model with a notion of pressure and chemical potential, which are equal in different phases. An example are binary and ternary liquid mixtures \cite{WoSc2006pre,IdLS2009pre} and related wetting phase transitions \cite{WoSc2004jpm,BEIM2009rmp}. An extension to three-dimensional systems is straight forward but still seems cumbersome due to limited computer power and the large number of possible phases, even for a binary system.

Note that continuation techniques may also be applied to deterministic lattice models or stochastic models \cite{ThLS2016pa,WTAL2020pre}. This implies that it should in the future also be possible to apply the proposed method to analyse the phase behaviour for lattice DFT and Monte Carlo models as studied, e.g., in \cite{ChCA2015jcp, BTAH2017jcp}.

%1d - 9 phases, 8 triple points

%2d - 21 phases, 24 triple points
%(2l+ 2hc+1sc)^2 possible, but su, sd, us, ud do not appear
% squares excluded

\acknowledgments

All authors thank the Center of Nonlinear Science (CeNoS) of the University of M{\"u}nster for recent support of the author's collaboration.  UT acknowledges support by the doctoral school ``Active living fluids’’ funded by the German French University (Grant No. CDFA-01-14); AJA acknowledges support by the EPSRC under grant No.~EP/P015689/1; We thank Tobias Frohoff-H\"ulsmann for frequent discussions on pde2path and PFC models, and Hannes Uecker for continued support of our continuation endeavours.

\clearpage
\bibliographystyle{abbrv}
\bibliography{phase_coexistence}

\begin{thebibliography}{10}

\bibitem{ARRS2019pre}
A.~J. Archer, D.~J. Ratliff, A.~M. Rucklidge, and P.~Subramanian.
\newblock Deriving phase field crystal theory from dynamical density functional
  theory: Consequences of the approximations.
\newblock {\em Phys. Rev. E}, 100:022140, Aug 2019.

\bibitem{AWTK2014pre}
A.~J. Archer, M.~C. Walters, U.~Thiele, and E.~Knobloch.
\newblock Solidification in soft-core fluids: {D}isordered solids from fast
  solidification fronts.
\newblock {\em Phys. Rev. E}, 90:042404, Oct 2014.

\bibitem{BEIM2009rmp}
D.~Bonn, J.~Eggers, J.~Indekeu, J.~Meunier, and E.~Rolley.
\newblock Wetting and spreading.
\newblock {\em Rev. Mod. Phys.}, 81:739--805, 2009.

\bibitem{BTAH2017jcp}
O.~Buller, W.~Tewes, A.~J. Archer, A.~Heuer, U.~Thiele, and S.~Gurevich.
\newblock Nudged elastic band calculation of the binding potential for liquids
  at interfaces.
\newblock {\em J. Chem. Phys.}, 147:094704, 2017.

\bibitem{BuKn2006pre}
J.~Burke and E.~Knobloch.
\newblock Localized states in the generalized {S}wift-{H}ohenberg equation.
\newblock {\em Phys. Rev. E}, 73:056211, May 2006.

\bibitem{BuKn2007c}
J.~Burke and E.~Knobloch.
\newblock Homoclinic snaking: structure and stability.
\newblock {\em Chaos}, 17:037102, 2007.

\bibitem{ChCA2015jcp}
B.~Chacko, C.~Chalmers, and A.~J. Archer.
\newblock Two-dimensional colloidal fluids exhibiting pattern formation.
\newblock {\em J. Chem. Phys.}, 143:244904, 2015.

\bibitem{ChGD2009pre}
P.~Y. Chan, N.~Goldenfeld, and J.~Dantzig.
\newblock Molecular dynamics on diffusive time scales from the
  phase-field-crystal equation.
\newblock {\em Phys. Rev. E}, 79:035701, 2009.

\bibitem{CrHo1993rmp}
M.~C. Cross and P.~C. Hohenberg.
\newblock Pattern formation out of equilibrium.
\newblock {\em Rev. Mod. Phys.}, 65:851--1112, 1993.

\bibitem{DWCD2014ccp}
H.~A. Dijkstra, F.~W. Wubs, A.~K. Cliffe, E.~Doedel, I.~F. Dragomirescu,
  B.~Eckhardt, A.~Y. Gelfgat, A.~Hazel, V.~Lucarini, A.~G. Salinger, E.~T.
  Phipps, J.~Sanchez-Umbria, H.~Schuttelaars, L.~S. Tuckerman, and U.~Thiele.
\newblock Numerical bifurcation methods and their application to fluid
  dynamics: {A}nalysis beyond simulation.
\newblock {\em Commun. Comput. Phys.}, 15:1--45, 2014.

\bibitem{DoKK1991ijbc}
E.~Doedel, H.~B. Keller, and J.~P. Kernevez.
\newblock Numerical analysis and control of bifurcation problems {(I)
  B}ifurcation in finite dimensions.
\newblock {\em Int. J. Bifurcation Chaos}, 1:493--520, 1991.

\bibitem{DoedelOldeman2009}
E.~J. Doedel and B.~E. Oldeman.
\newblock {\em AUTO07p: Continuation and bifurcation software for ordinary
  differential equations}.
\newblock Concordia University, Montreal, 2009.

\bibitem{ElGr2004pre}
K.~R. Elder and M.~Grant.
\newblock Modeling elastic and plastic deformations in nonequilibrium
  processing using phase field crystals.
\newblock {\em Phys. Rev. E}, 70:051605, Nov 2004.

\bibitem{ELWG2012ap}
H.~Emmerich, H.~Löwen, R.~Wittkowski, T.~Gruhn, G.~I. Tóth, G.~Tegze, and
  L.~Gránásy.
\newblock Phase-field-crystal models for condensed matter dynamics on atomic
  length and diffusive time scales: an overview.
\newblock {\em Adv. Phys.}, 61(6):665--743, 2012.

\bibitem{EGUW2019springer}
S.~Engelnkemper, S.~V. Gurevich, H.~Uecker, D.~Wetzel, and U.~Thiele.
\newblock Continuation for thin film hydrodynamics and related scalar problems.
\newblock In A.~Gelfgat, editor, {\em Computational Modeling of Bifurcations
  and Instabilities in Fluid Mechanics}, Computational Methods in Applied
  Sciences, vol 50, pages 459--501. Springer, 2019.

\bibitem{Glas2010sjam}
K.~B. Glasner.
\newblock Spatially localized structures in diblock copolymer mixtures.
\newblock {\em SIAM J. Appl. Math.}, 70:2045--2074, 2010.

\bibitem{HeBD2019pra}
V.~Heinonen, K.~J. Burns, and J.~Dunkel.
\newblock Quantum hydrodynamics for supersolid crystals and quasicrystals.
\newblock {\em Phys. Rev. A}, 99:063621, Jun 2019.

\bibitem{HiKA2009c}
Y.~Hirose, S.~Komura, and D.~Andelman.
\newblock Coupled modulated bilayers: A phenomenological model.
\newblock {\em ChemPhysChem}, 10(16):2839--2846, 2009.

\bibitem{Hoyle2006}
R.~B. Hoyle.
\newblock {\em Pattern formation -- An introduction to methods}.
\newblock Cambridge University Press, 2006.

\bibitem{IdLS2009pre}
T.~Idema, J.~van Leeuwen, and C.~Storm.
\newblock Phase coexistence and line tension in ternary lipid systems.
\newblock {\em Phys. Rev. E}, 80:041924, 2009.

\bibitem{Knob2016ijam}
E.~Knobloch.
\newblock Localized structures and front propagation in systems with a
  conservation law.
\newblock {\em IMA J. Appl. Math.}, 81:457--487, 2016.

\bibitem{KrauskopfOsingaGalan-Vioque2007}
B.~Krauskopf, H.~M. Osinga, and J.~Galan-Vioque, editors.
\newblock {\em Numerical Continuation Methods for Dynamical Systems}.
\newblock Springer, Dordrecht, 2007.

\bibitem{ophaus2018resting}
L.~Ophaus, S.~V. Gurevich, and U.~Thiele.
\newblock Resting and traveling localized states in an active
  phase-field-crystal model.
\newblock {\em Phys. Rev. E}, 98(2):022608, 2018.

\bibitem{RATK2012pre}
M.~J. Robbins, A.~J. Archer, U.~Thiele, and E.~Knobloch.
\newblock Modelling fluids and crystals using a two-component modified phase
  field crystal model.
\newblock {\em Phys. Rev. E}, 85:061408, 2012.

\bibitem{roth2010fundamental}
R.~Roth.
\newblock Fundamental measure theory for hard-sphere mixtures: a review.
\newblock {\em J. Phys.: Condens. Matter}, 22(6):063102, 2010.

\bibitem{scacchi2020quasicrystal}
A.~Scacchi, W.~R.~C. Somerville, D.~M.~A. Buzza, and A.~J. Archer.
\newblock Quasicrystal formation in binary soft matter mixtures.
\newblock {\em Phys. Rev. Research}, 2(3):032043, 2020.

\bibitem{SSAH2018jpm}
W.~R.~C. Somerville, J.~L. Stokes, A.~M. Adawi, T.~S. Horozov, A.~J. Archer,
  and D.~M.~A. Buzza.
\newblock Density functional theory for the crystallization of two-dimensional
  dipolar colloidal alloys.
\newblock {\em J. Phys.: Condens. Matter}, 30:405102, 2018.

\bibitem{TDME2019prm}
D.~Taha, S.~R. Dlamini, S.~K. Mkhonta, K.~R. Elder, and Z.-F. Huang.
\newblock Phase ordering, transformation, and grain growth of two-dimensional
  binary colloidal crystals: A phase field crystal modeling.
\newblock {\em Phys. Rev. Materials}, 3:095603, Sep 2019.

\bibitem{TARG2013pre}
U.~Thiele, A.~J. Archer, M.~J. Robbins, H.~Gomez, and E.~Knobloch.
\newblock Localized states in the conserved {S}wift-{H}ohenberg equation with
  cubic nonlinearity.
\newblock {\em Phys. Rev. E}, 87:042915, Apr 2013.

\bibitem{TFEK2019njp}
U.~Thiele, T.~Frohoff-Hülsmann, S.~Engelnkemper, E.~Knobloch, and A.~J.
  Archer.
\newblock First order phase transitions and the thermodynamic limit.
\newblock {\em New J. Phys.}, 21(12):123021, dec 2019.

\bibitem{ThLS2016pa}
S.~A. Thomas, D.~J.~B. Lloyd, and A.~C. Skeldon.
\newblock Equation-free analysis of agent-based models and systematic parameter
  determination.
\newblock {\em Physica A}, 464:27--53, 2016.

\bibitem{UeWR2014nmtmaa}
H.~Uecker, D.~Wetzel, and J.~D.~M. Rademacher.
\newblock pde2path - a matlab package for continuation and bifurcation in 2d
  elliptic systems.
\newblock {\em Numerical Mathematics: Theory, Methods and Applications},
  7(1):58--106.

\bibitem{WeKZ2013jcp}
V.~Weith, A.~Krekhov, and W.~Zimmermann.
\newblock Stability and orientation of lamellae in diblock copolymer films.
\newblock {\em J. Chem. Phys.}, 139:054908, 2013.

\bibitem{WTAL2020pre}
C.~Willers, U.~Thiele, A.~J. Archer, D.~J.~B. Lloyd, and O.~Kamps.
\newblock Adaptive stochastic continuation with a modified lifting procedure
  applied to complex systems.
\newblock {\em Phys. Rev. E}, 2020.
\newblock (at press).

\bibitem{WoKA2015pre}
J.~Wolff, S.~Komura, and D.~Andelman.
\newblock Budding of domains in mixed bilayer membranes.
\newblock {\em Phys. Rev. E}, 91:012708, 2015.

\bibitem{WoSc2004jpm}
D.~Woywod and M.~Schoen.
\newblock The wetting of planar solid surfaces by symmetric binary mixtures
  near bulk gas-liquid coexistence.
\newblock {\em J. Phys.: Condens. Matter}, 16:4761--4783, 2004.

\bibitem{WoSc2006pre}
D.~Woywod and M.~Schoen.
\newblock Topography of phase diagrams in binary fluid mixtures: a mean-field
  lattice density functional study.
\newblock {\em Phys. Rev. E}, 73:011201, 2006.

\end{thebibliography}
\end{document}